 \documentclass[sigconf]{acmart}

\pdfoutput=1
\usepackage{wrapfig}
\usepackage{colortbl}
\usepackage{tabularray}
\usepackage{color}
\usepackage[normalem]{ulem}
\usepackage{adjustbox}

\usepackage{amsthm} 

\usepackage{color,soul}
\usepackage{graphics}
\usepackage{hyperref}
\usepackage{url}
\usepackage{lipsum}
\usepackage{tikz}
\usepackage{enumitem}	
\usepackage{listings} 
\usepackage{booktabs}	
\usepackage{lipsum}
\usepackage{subcaption}
\usepackage{capt-of}	
\usepackage{diagbox}
\usepackage{multirow}
\usepackage{todonotes}  
\definecolor{refkey}{rgb}{249,158,26}
\definecolor{labelkey}{rgb}{0,0,1}
\definecolor{airforceblue}{rgb}{0.36, 0.54, 0.66}
\definecolor{applegreen}{rgb}{0.55, 0.71, 0.0}

\usepackage{mathrsfs}	
\usepackage{amsfonts}
\usepackage{algorithm,amsmath,algpseudocode} 
\usepackage{boldline}					

\usepackage{comment}
\usepackage{multirow}

\definecolor{frenzyorange}{RGB}{249, 158, 26}
\newcommand*\circled[1]{\tikz[baseline=(char.base)]{
		\node[shape=circle,fill=black,draw,text=white,inner sep=1pt] (char) {#1};}}

\newcommand*\whiteCircled[1]{\tikz[baseline=(char.base)]{
		\node[shape=circle,draw=black,inner sep=1pt] (char) {#1};}}
		
\renewcommand{\paragraph}[1]{\vskip 3pt\noindent\textbf{#1 }}	 

  {\begin{list}{$\bullet$}%
     {\setlength{\parsep}{0pt}%
      \setlength{\topsep}{0pt}%
      \setlength{\itemsep}{2pt}}}%
  {\end{list}}
\newcommand\Noted[1]{} 

\newcommand\xzlNote[1]{\sethlcolor{yellow} \hl{#1}} 
\newcommand\gray[1]{\textcolor{gray}{#1}} 

\definecolor{darkblue}{rgb}{0.0, 0.0, 0.55}
\definecolor{mygreen}{HTML}{ADFF2F}
\definecolor{mylightgray}{gray}{0.8}

\newenvironment{myitemize}%
  {\begin{itemize}
	[leftmargin=0cm,
		itemindent=.3cm,
		labelwidth=\itemindent,
		labelsep=0pt,
		parsep=1pt,
		topsep=1pt,
		itemsep=1pt,
		align=left]
  }%
  {\end{itemize}}    

  {\begin{enumerate}
	[leftmargin=.cm,itemindent=.5cm,labelwidth=\itemindent,
		labelsep=0pt,
		parsep=1pt,
		topsep=1pt,
		itemsep=3pt,
		align=left]
  }%
  {\end{enumerate}}    

\newcommand\sect[1]{Section~\ref{sec:#1}}	

 \newcommand{\sys}{SC}

\newcommand{\slurp}{SLURP-C}
\newcommand{\slurpMix}{SLURP-mix}

\makeatletter
\def\@copyrightspace{\relax}
\makeatother

\copyrightyear{2024}
\acmYear{2024}
\setcopyright{rightsretained}
\acmConference[MOBISYS '24]{The 22nd Annual International Conference on Mobile Systems, Applications and Services}{June 3--7, 2024}{Minato-ku, Tokyo, Japan}
\acmBooktitle{The 22nd Annual International Conference on Mobile Systems, Applications and Services (MOBISYS '24), June 3--7, 2024, Minato-ku, Tokyo, Japan}\acmDOI{10.1145/3643832.3661886}
\acmISBN{979-8-4007-0581-6/24/06}
\begin{abstract}

This paper addresses spoken language understanding (SLU) on microcontroller-like embedded devices, 
integrating on-device execution with cloud offloading in a novel fashion. 
We leverage temporal locality in the speech inputs to a device and reuse recent SLU inferences accordingly. 
Our idea is simple: let the device match incoming inputs against cached results, and only offload inputs not matched to any cached ones to the cloud for full inference. 
Realization of this idea, however, is non-trivial: 
the device needs to compare acoustic features in a robust yet low-cost way. 

To this end, we present SpeechCache (or \sys{}), a speech cache for tiny devices.
It matches speech inputs at two levels of representations: first by sequences of clustered raw sound units, then as sequences of phonemes. 
Working in tandem, the two representations offer complementary tradeoffs between cost and efficiency. 
To boost accuracy even further, our cache \textit{learns} to personalize: 
with the mismatched and then offloaded inputs, 
it continuously finetunes the device's feature extractors with the assistance of the cloud. 

We implement \sys{} on an off-the-shelf STM32 microcontroller. 
The complete implementation has a small memory footprint of 2 MB. 
Evaluated on challenging speech benchmarks, 
our system resolves 45\%--90\% of inputs on device, 
reducing the average latency by up to 80\% compared to offloading to popular cloud speech recognition services. 
The benefit brought by our proposed \sys{} is notable even in adversarial settings -- noisy environments, cold cache, or one device shared by a number of users.

\end{abstract}


\copyrightyear{2024}
\acmYear{2024}
\setcopyright{rightsretained}
\acmConference[MOBISYS '24]{The 22nd Annual International Conference on Mobile Systems, Applications and Services}{June 3--7, 2024}{Minato-ku, Tokyo, Japan}
\acmBooktitle{The 22nd Annual International Conference on Mobile Systems, Applications and Services (MOBISYS '24), June 3--7, 2024, Minato-ku, Tokyo, Japan}\acmDOI{10.1145/3643832.3661886}
\acmISBN{979-8-4007-0581-6/24/06}

\begin{document}

\title{Speech Understanding on Tiny Devices with A Learning Cache}



\author{Afsara Benazir}
\authornote{Both authors contributed equally to the paper}

\affiliation{%
	\institution{University of Virginia}
	\city{}
	\state{}
	\country{}
}
\email{hys4qm@virginia.edu}
\author{Zhiming Xu}
\authornotemark[1]
\affiliation{%
	\institution{University of Virginia}
	\city{}
	\state{}
	\country{}
}
\email{zx2rw@virginia.edu}

\author{Felix Xiaozhu Lin}
\affiliation{%
	\institution{University of Virginia}
	\city{}
	\state{}
	\country{}
}
\email{felixlin@virginia.edu}


\date{}

\maketitle

\vspace{-2mm}
\section{Introduction}
\label{sec:intro}
Speech is a pervasive human-computer interface. 
It is in particular suitable to small embedded devices where form factors are limited and hands-free interactions are preferred. 
Examples include voice user interface, smart home gadgets (such as smart lights/speakers/thermostat), 
and fitness trackers. 
The tasks that underpin speech interfaces are spoken language understanding (SLU), 
which translates a spoken utterance in the form of an audio wave to a structured semantic label, e.g. 
\textit{``Turn the temperature in the bedroom up''} is mapped to  \textit{\{intent: ``change\_temperature'', slots: [\{action: increase, object: heat, location: bedroom\}]}. 
By 2023, 4.2 billion devices are estimated to have voice input capabilities \cite{stat}. 



To deploy SLU on small embedded devices\footnote{Referred to as ``devices'' for simplicity; to be defined in \sect{motiv}}, 
neither on-device execution nor offloading is ideal: 

(1) On-device execution is constrained by resources. 
Satisfactory SLU accuracy often requires a deep neural network, typically an encoder-decoder architecture with transformers \cite{denisov2020pretrained, chung2020splat, haghani2018audio}.
Their memory footprint (hundreds of megabytes) far exceeds the memory on tiny devices which is in the order of a few megabytes ~\cite{zhang2017hello}. 
While recent research has tailored SLU to small devices~\cite{coucke2018snips, arik2017convolutional}, the resultant models are limited to  predefined, simple commands such as ``Lights up''; 
the models fall short of handling longer and more complex speeches commonly seen in daily life \cite{long2023ai}. 

(2) Offloading inputs to the cloud incur higher delays and costs. 
The delays consist of waking up network interfaces, establishing connections, and transmitting data.
Users are well-known to be sensitive to delays in speech interactions~\cite{kohrs2016delays, porcheron2018voice}. 
The costs of cloud inference is also a rising concern, 
as shared by both research communities~\cite{vipperla2020learning} and industry stakeholders~\cite{qualcomm2023hybridcost, microsoft2023azurehybridbenefit, qualcomm-report}.
As a notable example, Qualcomm estimates that each machine learning (ML) query costs 10x higher compared to conventional cloud API \cite{qualcomm-report}.



To this end, this paper sets to integrate on-device execution (for reduced delay/cost) with cloud offloading (for well-established accuracy) in a novel way. 

\paragraph{Insight: locality in spoken inputs}
Our key observation is the temporal locality in voice inputs:
for a device, the received voice commands are likely uttered by a small number of users; 
most voice inputs are recurring with nearly identical transcripts. 
See \S \ref{sec:motiv:locality} for detailed evidence. 

The temporal locality implies a new opportunity on system efficiency: 
the device may match any new input against cached ones and directly emit user intents in case of high similarity. 
The rationale is that the device could resolve a significant fraction of inputs without offloading, 
reducing latency while still retaining accuracy across all inputs. 

While the idea seems simple, to realize it we are met with daunting challenges. For one thing, raw speech signals exhibit strong acoustic variability as induced by disturbance and noises \cite{lim2021multispeaker}. 
For the other, even utterances of the \textit{same} transcript, by the \textit{same} speaker, will vary notably in the acoustic domain, which renders straightforward waveform comparisons ineffective. 
The drastic difference can be seen in \autoref{fig:speech-comp}. 

\paragraph{Match at which level of representations?}
The choice of voice features requires a careful examination and thoughtful tradeoff. 
On one hand, to be robust against variations, our system should match utterances represented by higher-level speech features. 
On the other hand, extracting higher level features is beyond the capacity of small feature extractors that can fit in the devices. 
In principle, our chosen speech representations should  generalize beyond individual utterance instances; 
yet, they should not \textit{over-generalize}, e.g., extracting text from utterances becomes an overkill for matching repeated utterances. 


\begin{figure}[t]
	\includegraphics[width=0.48\textwidth]{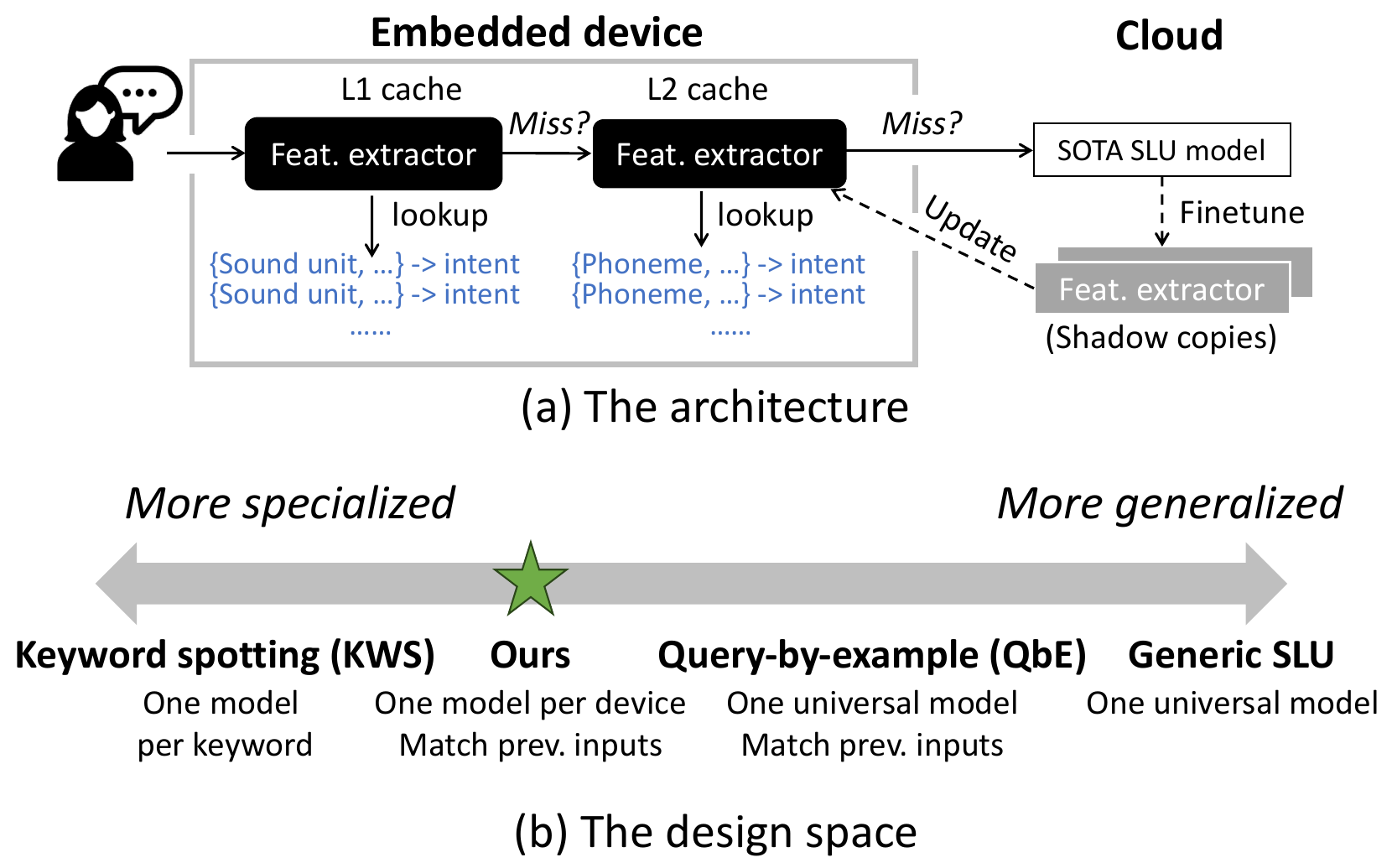}
	\caption{Our system in the design space of speech understanding systems.}
	\label{fig:overview}
		\vspace{-7mm}
\end{figure}

\paragraph{How to ensure accuracy?}
Lightweight feature extractors often output noisy features. 
To achieve competitive match accuracy, our ideas are twofold: 
(1) \textit{specialization}: 
instantiating multiple versions of feature extractors 
for one device and for utterances of similar lengths; 
(2) \textit{learning from offloading}: 
using the SLU results returned by the cloud as a supervision signal, 
continuously tuning the on-device feature extractors. 
Combined, the two ideas specialize a device's extractors to the inputs received by the device \textit{in situ}.

\paragraph{Our system}
is a learning cache for SLU on embedded devices. 
As shown in \autoref{fig:overview}(a), \sys{} entails two major designs: 
(1) it caches recent offload results: 
input utterances as represented by low or middle level acoustic features; 
(2) it uses the recent offload results to finetune the device's feature extractors for specialization. 
\sys{} matches new inputs first by clustered raw frequency vectors (referred to as L1, a low level representation) and in case of L1 mismatch, by CTC (Connectionist Temporal Classification) 
loss \cite{graves2006connectionist} over phoneme sequences (referred to as L2, a middle level representation). 
Notably, \sys{}'s feature extractors are streaming, computing on utterance segments as they are ingested, 
which hides much of the computation delay behind the IO.



As shown Figure~\ref{fig:overview}(b), \sys{} takes a new point in the design space of speech understanding. 
\sys{}'s choice of acoustic features sets it apart from work that matches voice commands by transcripts~\cite{9355621}; 
\sys{}'s learning capability sets it apart from query-by-example (QbE) \cite{hazen2009query, chen2015query, lugosch2018donut, kim2019query} and keyword spotting (KWS) \cite{arik2017convolutional, mittermaier2020small} which cannot personalize online. 
To ensure reproducibility, we release our training and inference code along with sample models at
the following URL: \href{https://github.com/afsara-ben/SpeechCache}{https://github.com/afsara-ben/SpeechCache}





\paragraph{Results} 
We implemented \sys{} on a low-cost MCU and evaluated it on 75K+ voice inputs from 210 users. 
Compared to offloading all inputs, \sys{} resolves 45\% -- 90\% of inputs with recurring transcripts on device, reducing the average latency by up to 81\% (from near 1 second down to around 150 ms). 
Meanwhile, it retains accuracy comparable to that of the state-of-the-art models.
Even in challenging situations such as scarce cached inputs or multi-tenant usage of the device, 
\sys{} still reduces the latency by 3\% -- 34\% compared to offloading to the cloud.
\sys{} has a low memory footprint of less than 2 MB (model and cache combined), 
which is suitable to low-cost embedded devices. 

\paragraph{Contributions}
    Towards deep speech understanding on small embedded devices: 
    
\begin{myitemize}
    \item 
    We identify a fresh opportunity: temporal locality in voice inputs, 
    and leverage it through on-device caching. 
    
    
    \item 
    We propose two novel designs: 
    matching utterances at hierarchical levels of acoustic features; 
    personalizing the feature extractors, 
   	which is done via finetuning the extractors with offload results.

	\item We present a suite of optimizations which are vital to performance, including an ensemble of models for various input lengths, dynamic thresholds, and model compression.

\end{myitemize}

\section{Motivations}
\label{sec:motiv}

\subsection{Our system model}
\label{sec:motiv:m}

We target embedded devices as commonly seen in smart home and wearable gadgets. 
Examples include Arm Cortex-M microcontrollers clocked at less than a few hundred MHz, with less than a few MBs of RAM. 
We will refer to them as ``devices'' for short. 

Following prior work, 
we assume that a device continuously detects the presence of human voices with low overhead, 
and activates SLU upon detection~\cite{zhang2017hello}. 
We assume that the device has network connectivity -- to a cloud service or a nearby hub, 
which we refer to as ``cloud'' for short. 
The cloud serves a state-of-the-art (SOTA) SLU model. 
The resource needed by such an SLU model (memory in 100s of MB) far exceeds that of a device. 

\paragraph{The goal and target benefits} We set to make the devices capable of executing SLU with near SOTA accuracy, while processing a significant fraction of inputs locally. 
Our primary benefits are (1) lower cloud cost by reducing the invocations to its ML service; 
(2) lower inference latency and thus better user experience. 
Beyond that, our system may strengthen privacy and security for applications that allow occasional offloading despite the risk of data breach.

\subsection{A primer: SLU on tiny devices}
\label{sec:slu}






The conventional SLU architecture follows a pipelined approach with separate encoder-decoder or transformer based Automatic Speech Recognition (ASR) module followed by a large language model (LM) \cite{huang2023leveraging, zhu2017encoder, qin2021survey, qin2019stack, chen2019bert}. This pipeline can be compute expensive, error-prone \cite{serdyuk2018towards, radfar2021fans} and is non-personalized as the modules are trained separately. Moreover, prosody is overlooked in ASR transcription limiting complete contexual understanding. End-to-end SLU (E2E SLU) is more compact and skips the need for separate modules \cite{serdyuk2018towards}.  Given an input sequence $X=\{x_1, ...., x_t\}$, the encoder $E$, produces a fixed-size embedding $H=\{h_1, ..., h_t \}$, for each element in $X$; $H=E(X)$. The decoder $D$ generates the output sequence $Y=\{y_1, ..., y_t\}$ based on the embedding $H$; $Y=D(H)$.

\begin{figure}[t]
    \centering
    \includegraphics[width=\columnwidth]{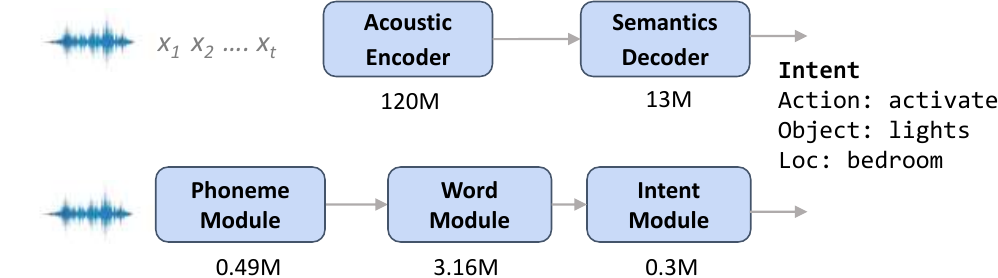}
\caption{Typical models and parameter sizes for SLU. (Top) a large, conventional model \cite{huang2023leveraging}; (Bottom) an end-to-end model compressed for MCU, 
which often can only understand simple commands}
    \label{fig:SLUmodel}
\vspace{-5mm}
\end{figure}

A large speech recognition model \cite{huang2023leveraging}
can resolve rich, complex commands that resemble daily conversations (e.g. '\textit{please check and tell me about the reminders i have placed today}') but is beyond the compute and memory resources of a simple MCU. It is illustrated in \autoref{fig:SLUmodel}.
SLU under resource constraints is feasible only with certain restricted input types: utterance is short and is a predefined keyword (“yes”, “stop", "go" etc.) \cite{arik2017convolutional, mittermaier2020small} or few word command ("play music", "lights on") \cite{lugosch2019speech}. 
In \cite{lugosch2019speech} a three slot tuple\textit{ \{action (A), object (O), location (L)\}} is defined as an intent. Intent classification (IC) is a semantic utterance classification problem \cite{liu2016joint} -- $P(A, O, L \mid D) = P(A \mid D) \cdot P(O \mid D)  \cdot  P(L \mid D)$, given a sequence of acoustic feature $D$ from a given utterance \cite{palogiannidi2020end}.

\paragraph{Cost breakdown}
Typical E2E large SLU model in \autoref{fig:SLUmodel} (a) although being highly accurate, is impossible to deploy on MCU.
To understand caching opportunities and the potential reduction in overhead, 
we study a small SLU model \cite{lugosch2019speech} in \autoref{fig:SLUmodel} (Bottom).
Note that the model was designed with efficiency in mind, not accuracy. Its capacity is much lower than \sys{} (ours) as discussed in \S \ref{sec:end-to-end}.
Our observations are that: 
(1) the early stages require much less compute and memory than later stages. 
These stages are for extracting lower level acoustic features (sound unit, phoneme). 
The later stages, especially word embeddings, require memory that far exceeds embedded devices. 
(2) executing the early stages take much less time ($\sim$96ms) than offloading one second of input to the cloud
($\sim$ 300ms).
Note that the choice of offloading audio waveform ($\sim$10s KB per input) or intermediate features (e.g. phonemes at 160 bytes per input second) does not affect the offloading latency much; 
the delay is primarily contributed by the network round trips (RTT) and the device's network interface power management (wakeup or duty cycling) \cite{mtibaa2013towards}.

\paragraph{Applicability}
Our system targets smart devices, typically interface devices for controlling smart homes, used by one/several users and seeing recurring queries or commands. Any Voice User Interface (VUI), smart wearables, smart home control and gadgets such as thermostats, light control fall under this category. Additionally, our system has use cases for human robot interaction and smart navigation.

\subsection{Motivation: locality in spoken inputs}
\label{sec:motiv:locality}

Such input locality is pervasive, for the following reasons. 

\noindent
\textbf{In many settings, a smart device serves no more than several users.} 
For instance, 
a wearable serves its wearer exclusively; 
home devices in a bedroom serve family members that use the room~\cite{bentley2018understanding};
smartspace devices such as activity trackers serve one or a few workers in close proximity. 

\noindent
\textbf{Speech inputs received by a device often follow a small set of transcripts.} 
We attribute the causes to two sources of locality. 
(a) \textit{Intent locality}. 
Studies show that, despite rich features of home assistants, 
most invocations are for a small set of daily routines, such as querying schedules, checking weather, and IoT controls \cite{ammari2019music}.
The users rarely change their usage patterns, and the majority of sessions (77\%) involves only one or two domains~\cite{bentley2018understanding}.
(b) \textit{Transcript locality}. 
To express the same intent, 
a user is likely to utter the same transcript repeatedly, e.g. ``what time is it'' for querying the time. 
Study shows that humans naturally follow linguistic/social conventions to facilitate understanding among individuals~\cite{mctear2016conversational}. 
The \textit{idiolect study} also demonstrates that a user often has her own preferred choice of words and sentences~\cite{coulthard2004author}. 
Such human-to-human conventions are reinforced in the human-computer interactions; users are motivated to use recurring transcripts 
to facilitate machine understanding. 

We recognize that the above observations may not apply to certain devices, e.g. a public kiosk serving diverse users/inputs.
For them, our system will not fail but will see diminishing benefits, 
as \S \ref{sec:eval} will show.






%

\subsection{Design implications}

The locality motivates us to cache recently spoken inputs, 
for which we make three design choices. 

\begin{figure}[h]
    \centering
    \includegraphics[width=\columnwidth]{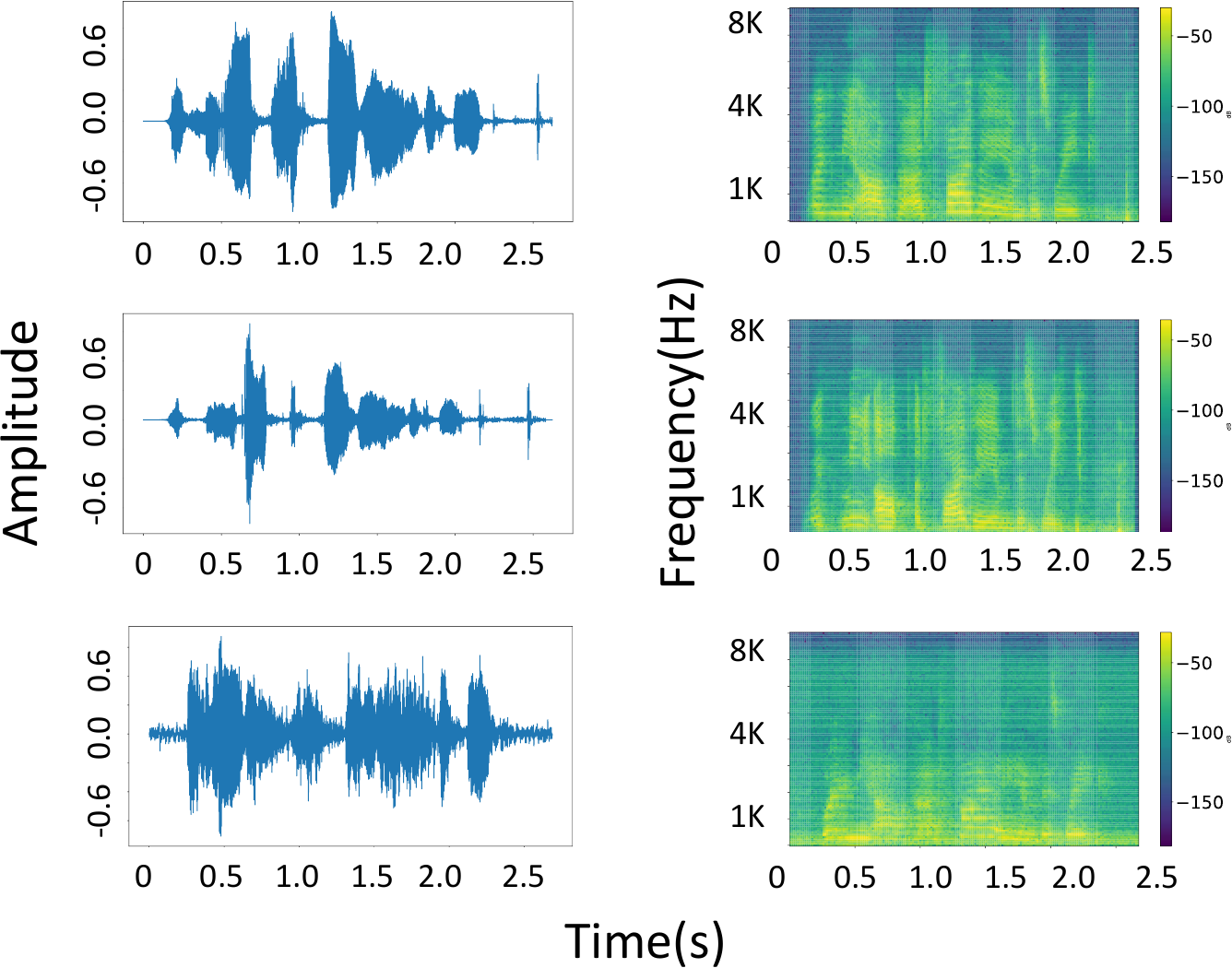}
    \caption{Raw speech waveforms often exhibit strong variations. 
    Three waveforms of the same sentence (``wake me up at 5am this week''), uttered by the same speaker, from SLURP are shown. First two are recorded in far field conditions while the third is recorded in close range (headset).}
    \label{fig:speech-comp}

\end{figure}

\paragraph{Choosing cache representations}
Our top challenge is robust match of utterances at low cost. 
The solution hinges on the representations of cached inputs. 
(1) While low level acoustic representations require smaller ML models and less computations, 
the representations can be brittle. 
This is further exacerbated by background noises and the speaker distance. 
Figure~\ref{fig:speech-comp} compares three spoken instances of the same transcript, by the same user back to back. 
These representations not only appear visually different, 
but also fail common metrics for sequence matching: 
euclidean similarity, cosine similarity, and Levenshtein distance. 
(2) By contrast, higher level representations such as words are more robust; 
yet, the required on-device memory and computation would defeat our goal of efficiency.

Considering these tradeoffs, we choose to represent cache inputs as sequences of acoustic or phoneme features. 

\paragraph{Specialization}
The cache adapts to each device and its users, and more aggressively, each recent transcript. 
This allows \sys{} to specialize its model parameters (mapping from raw waveforms to acoustic representations) and cache representations. 
This drives the accuracy high and allows simple model structures.  

\paragraph{Learning online}
The device can receive online supervision from the cloud model in situ, 
using the cloud output to finetune its local models for specialization. 
This sets our design apart from prior embedded speech systems \cite{zhang2017hello, lugosch2018donut, lugosch2019speech, chen2014small}
which lack such supervision; 
these systems run generic models frozen at \textit{development time} and compromise on accuracy. 

\section{\sys{} Overview}
\label{sec:design}

\paragraph{Architecture}
\sys{} consists of a device runtime that extracts input features and performs cache lookup;
a cloud runtime, which resolves offloaded inputs with a SOTA SLU model; 
the cloud finetunes the feature extractors for the device. 

On a device, the cache comprises two levels: 
level one (L1) represents an input as a sequence of sound units and level 2 (L2) represents an input as a sequence of phonemes. 
The two levels run ML models as their feature extractors. This is illustrated in \autoref{fig:overview}. 
The maximum number of cached inputs is user configurable, 
which we expect to be no more than one hundred based on workload studies \cite{bentley2018understanding, garg2020he, sciutohey}.


To extract the features at two levels, \sys{} runs two lightweight acoustic models: (1) SincNet \cite{ravanelli2018speaker} and 1D convolutions, followed by frame-level frequency discretization and (2) GRUs with linear classification. 

We further apply a series of optimizations (discussed in \S \ref{para:audio-bucket}): augmenting input data for finetuning, 
grouping utterances by lengths and running separate feature extractors, 
and dynamic thresholds for similarity comparisons. 

\paragraph{Operations}
When deployed, 
\sys{} installs generic, pretrained feature extractors to a device. 
It uploads a shadow copy of the feature extractors to the cloud. 

During operation, 
any new input goes through the L1 cache and, in case of L1 mismatch, the L2 cache. 
Each cache level generates input representations and compares them against existing entries. 
In case of both L1 and L2 mismatch, 
\sys{} uploads the input -- as a raw waveform -- to the cloud. 
The size of a waveform contributes little to the upload delay: 
a typical input lasting 3 seconds is no more than tens of KB, 
for which the upload delay is dominated by a network round trip \cite{mtibaa2013towards}.
After processing an offloaded input, 
the cloud sends back the resultant intent as well as new L1/L2 entries (as sequence of sound units and phonemes, respectively), 
which the device installs.
Note that upon L1 or L2 cache hit, the cloud is not invoked; no new cache entries are installed. 

In processing offloaded inputs,
the cloud also finetunes its shadow copy of the feature extractor. 
Every $N$ offloaded inputs, the cloud sends the shadow copy back to the device, replacing its local feature extractors. 
Empirically, we find $N=100$ as a good balance between model freshness and download frequencies. 
The model download is as fast as 1--2 seconds, as a single model size is < 2 MBs before compression and 0.62 MBs \textit{after} compression. See \S \ref{overhead-analysis} for details on memory footprint.
For finetuning details, see \S \ref{sec:L1} and \S \ref{sec:L2}.


\sys{} caches multiple entries per transcript as utterance for the same transcript varies; evident in \autoref{fig:speech-comp}.  Cache entry per utterance is updated at runtime in absence of cache hit.  

The overall latency for \sys{} is
calculated taking into account the percentage of inputs that
are offloaded. Since \sys{} is streaming, only the last 10 frames or 0.25s of audio processing adds to the latency overhead. The choice of segment/step size is consistent with prior work \cite{mhiri2020low}. The filter rate is the percent number of inputs
that are processed on-device. A higher filter rate thus implies
a lower latency as most processing would be done locally. 

\paragraph{Cache space management}
Since an entry takes as few as 200 bytes (see Table \ref{tab:comp} for details), 
\sys{} keeps all entries in memory by default. 
We noticed that: a mid-range MCU of 2 MB memory can easily allocate 1\% of its flash for 50-60 cache entries, sufficient for covering most recent received utterances. 
After all free entries are used, the device invokes the well-known LRU policy \cite{johnson19942q} to evict victim entries. 
The device allows to cache multiple inputs that map to the same intent. 
It, however, prevents a small number of popular intents from dominating the whole space by capping the number of per intent entries. 



\paragraph{Cache warm up}
For a new device without prior inputs, \sys{} preloads the cache with example entries. 
These entries can be chosen for specific deployment, 
e.g. common voice commands in smart homes. 
In this warm up period, our cache resolves fewer inputs (i.e. lower hit rates) but still functions. 
As the cache encounters new commands and replaces the preloaded entries, its accuracy will ramp up to normal. 

\paragraph{Alternative designs rejected by us}
While it may be tempting to specialize for individual speakers (rather than devices as \sys{} does), we keep \sys{} simple without any need for speaker identification. 
However, \sys{} altogether caches inputs that may come from different speakers sharing a device.
We experimentally test the impact on the number of users per device in \S \ref{sec:end-to-end}. 

\sys{} matches intents with \textit{identical} transcripts.
It cannot match utterances of different transcripts implying same intent
(``Turn off the light'' cached, \sys{} still regards ``Turn the light off''  as a cache miss). 
Trivial fixes, e.g. raising the similarity threshold, fails because intents are sensitive to slight differences in transcripts (``Turn \textit{bed}room light on'' and ``Turn \textit{bath}room light on'' imply different intents).
The systematic fix is transitioning from acoustic representation to word embeddings, which however defeats \sys{}'s efficiency goal.
\section{Sound unit cache (L1)}
\label{sec:L1}

\begin{figure}[t]
	\centering
	\includegraphics[width=\columnwidth]{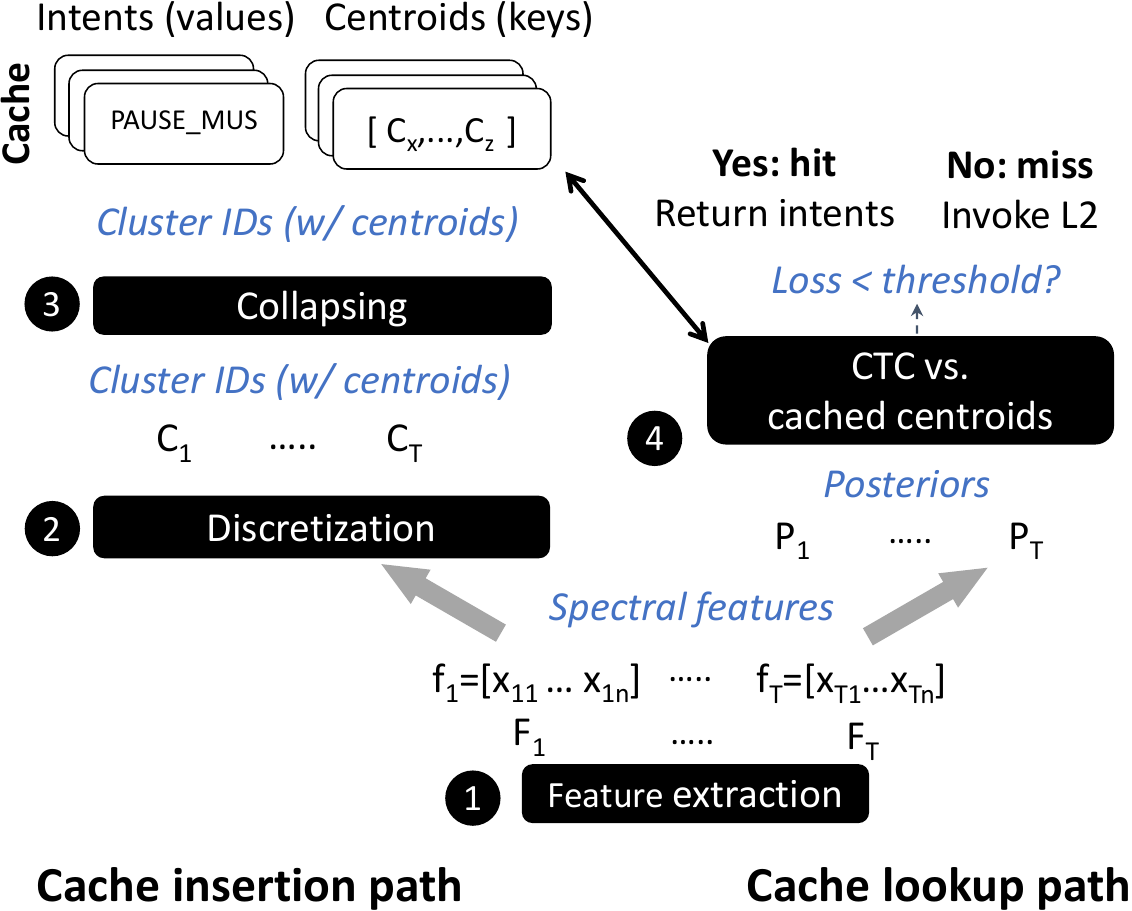}
%
%
 \caption{The design of sound unit cache (L1)}

	\label{fig:l1d}
\end{figure}


Our rationale for L1 is to absorb ``easy'' audio inputs (e.g. highly similar to some prior inputs). 
For this reason, we design L1 cache for low cost, and in exchange tolerate a lower recall
(i.e. L1 will miss a significant amount of true recurring inputs).

\subsection{Overview}

L1 compares low level spectral features of sound units as shown in \autoref{fig:l1d}. 

At runtime, given a waveform input $L$, the system \ul{queries} the cache by: extracting spectral features from $L$ (\circled{1}) and matching the frame-wise features against each cached key, which is a sequence of spectral feature centroids. 
For the matching, the system computes the CTC loss between the incoming features with the cached centroids (\circled{4}). 


In case of cache miss, the system \ul{updates} the cache by: 
discretizing the frame-wise features through clustering (\circled{2}) 
and further collapsing adjacent frames (\circled{3}). 
It saves the resultant sequence of centroids $\{C_i\}$ 
(which represent distinct sound units) alongside the cloud-supplied intent. 




\subsection{Designs}

We next elaborate on the major steps in \autoref{fig:l1d}.

\paragraph{\circled{1}~Feature extraction} 
L1 splits the input into $T$ audio frames $F_i  = \{F_1, ... , F_T\}$.
For each frame $F_i$, it extracts spectral features 
$f_i = \{x_1, x_2, ... , x_n \}$ which is a vector of continuous real numbers.

An audio frame is a short segment or window of consecutive audio samples having a fixed duration. A distinct sound constituting a single or multiple frame(s) is referred to as a sound unit or phone.
Converting audio frames to salient, spectral features is a standard preprocessing step of all modern SLU pipelines \cite{ravanelli2018speaker, mittermaier2020small}. It is also required by L2. Therefore, the extraction is free/independent to L1. It is of low cost, as done by a sequence of convolutional layers. 
Notably, we operate in the time domain: 
apply convolutions on a raw input waveform, instead of transforming the waveform to the frequency domain \cite{palaz2015analysis, dai2017very, ravanelli2018speaker}. 

\sys{} sends each frame to a SincNet layer \cite{ravanelli2018speaker} 
followed by two 1D convolutional layer. 
These layers build upon the high-dimensional features acquired in the initial SincNet layer. The output features are of lower abstraction.



Since the extraction is generic to human speech, \sys{} adopts frozen SincLayer and convolutional layers pretrained for ASR. 
For pretrained feature extractors, we use the extractor in phoneme module from Fluent \cite{lugosch2019speech}. Finetuning the frozen pretrained layers saw little benefit.

\paragraph{\circled{2}~Feature discretization}
\sys{} discretizes spectral feature vectors per frame through K-means clustering \cite{lloyd1982least}, as inspired by self-supervised speech models~\cite{coates2012learning, baevski2021unsupervised}. 
To do so it clusters vectors based on their euclidean distance from randomly initialized centroids. Each vector is assigned to its nearest centroid and the centroid positions are updated iteratively until convergence. With each iteration L1 learns more about the inherent characteristics present within the feature vector. 
After discretization, each frame $F_i$ is represented by a numerical ID $C_i$ for $i = 1 ... T$.

The extracted $T$ spectral feature vectors, $f = \{f_1, ...., f_T\}$ are continuous in nature but for our sequence matching task, we need discrete values. We opt for discretizing the frame features before sending them to the subsequent ML layers (as prediction targets). The rationale is that each discretized feature could represent a distinct sound (roughly), simplifying comparisons/matching of the sounds. 



Unique to \sys{}'s specialization principle, the clustering is per utterance instead of per transcript, so each utterance has its distinct sequence of centroids.
At runtime, for each seen transcript $t$, \sys{} augments the speech as described in \S \ref{sec:opt} and clusters the extracted spectral vectors corresponding to $t$, using K-means clustering.
Whenever the device encounters a new utterance for $t$, the sequence of centroids $C_i  = \{C_1, ... , C_T\}$ against $t$ are updated. The results are separate sets of $K$ centroids each for a distinct utterance of $t$.



We also experiment with $K$ centroids \textit{per device} instead of \textit{per utterance}. 
The performance is inferior to the prior approach because it is too coarse grained for varied transcripts. A set of $K$ centroids per device is inconsistent with the original transcript, hence we reject the idea. 


\paragraph{\circled{3}~Frame collapsing}
L1 collapses adjacent frames that have identical IDs (e.g. ``... 13 42 42 56 56 ...'' becomes ``... 13 42 56 ...''). The discretized sequence $C_i$ is converted to a sequence of sound units $C = \{C_x, C_y\cdots,C_z\}$ that the device caches against its respective intent. Coalescing adjacent IDs is important as it produces a sequence of sound units each of which may span multiple audio frames. 
\paragraph{\circled{4}~Sequence matching}
\sys{} does a ``soft'' match between a sequence of \textit{probabilistic} sound units against cached sequences. 
Due to the noisy nature of audios, 
``hard'' match between deterministic sound units would force \sys{} to emit one most likely sound unit per frame, which does not optimize for a most likely sequence match as a whole and is therefore susceptible to input disturbance. 
%


For each spectral feature $f_i$ in $F_i$ it generates a probability distribution $P_i$ over all discrete sound units. The distribution is calculated based on the distances to the cached centroids $\mathbf{C}=\{C_1,C_2,\cdots,C_j\}$ as: $\mathbf{d}_i=\left[\operatorname{d}(\mathbf{f}_i, \mathbf{c}_1),\operatorname{d}(\mathbf{f}_i, \mathbf{c}_2),\cdots,\operatorname{d}(\mathbf{f}_i, \mathbf{c}_j)\right].$ Since the closer a spectral feature is to a certain centroid, the more likely it belongs to that centroid, we utilize the inverse of the distance to measure the likelihood to match. Specifically, for time frame $i$, the inverse is calculated with $\mathbf{I}_i=\max(\mathbf{d}_i)-\mathbf{d}_i$. The predicted match to a specific cached centroid, $\mathbf{P}_i$, is obtained with $\mathbf{arg min}_j \mathbf{d}_i$.








Given the probabilistic $P_i$ and a cached sequence ($L_i$), L1 determines \underline{all} \textit{concrete} sequences of centroids (alignment) that would collapse to $L_i$. 
It computes the aggregated probabilities for a single alignment ($\pi$) as $ p(\pi|P) = \prod_{t=0}^{T-1} y_{\pi_t}^t$ where $y_{\pi_t}^t$ is  the probability of observing label $\pi_t$ at time $t$.

After that, \sys{} uses an objective function CTC loss $(l)$ \cite{graves2006connectionist}, a well known algorithm in speech recognition, 
to compute the aggregated probability of all possible input sequences (or valid alignments) in the inverse mapping $\tau^{-1}(l)$ that would collapse to a cached sequence: $L_i = p(l|P) = \sum_{\pi \in \tau^{-1} (l)} p(\pi|P)$



The process of finding the best alignment (sequence) is handled using a dynamic programming algorithm called the forward-backward search which has the lowest loss value. 
If the loss is below a pre-defined threshold X, \sys{} deems it as a cache hit and returns the associated intent. 
Intuitively, the threshold X should be normalized to the sequence length 
(e.g. tolerating higher loss for longer sequences). 

\subsection{Implementation details}
\label{l1-imp}
At a standard sampling rate of 16KHz \cite{Jurafsky2009}, we apply a hamming window of size 401 to $L$, creating a sequence of $T$ frames, each spanning 25 ms, or 401 audio samples.

In {\circled{1}} each 1D convolutional layer consists of 60 filters of length 5 and step size 1. To decrease the dimensionality of features, each layer undergoes a temporal 1D average max-pooling operation with a kernel size of 2 and a stride of 2. All hidden layers use leaky RELU activation function with a negative slope of 0.2 and batch normalization is applied before each activation layer to speed up training convergence. 

\paragraph{Hyperparameters} Through ablation study we find that for L1, a k-means cluster size of 70, euclidean edit distance, tolerance  $1e^{-4}$, a 3-layer convolutional layer having kernel size of (401, 5, 5) and stride (80, 1, 1) is optimal.

\section{Phoneme cache (L2)}
\label{sec:L2}
\begin{figure}[t]
		\centering
\includegraphics[width=0.5\textwidth]{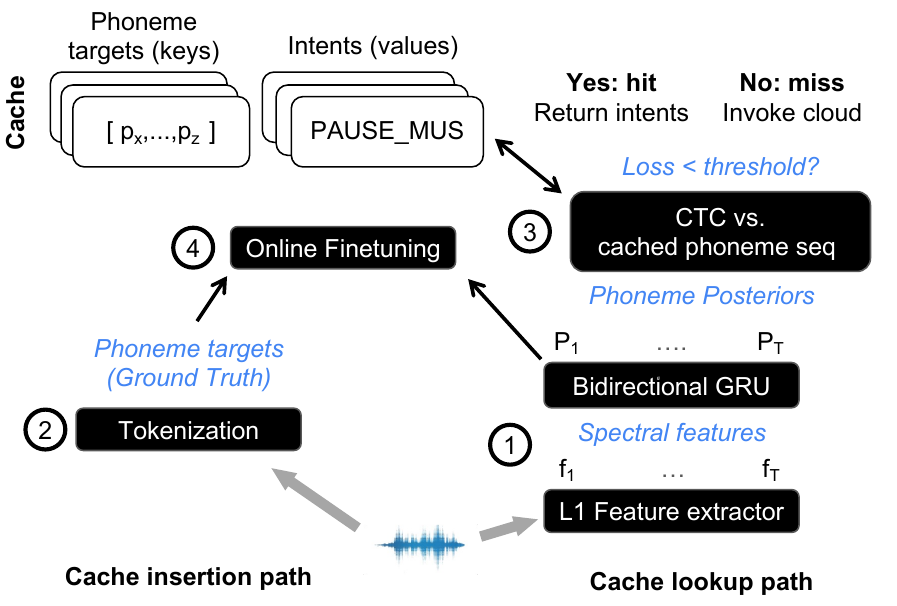}
	\caption{The design of phoneme cache (L2). 
 }
	\label{fig:L2-design}
		\vspace{-5mm}
\end{figure}

\subsection{Overview}
L2 matches inputs based on phonemes as shown in \autoref{fig:L2-design}.


At runtime given a waveform input $L$, L2 runs a feature extractor: the input is a sequence of spectral features (continuous) produced by L1 feature extractor in \autoref{fig:l1d}; 
the output is a sequence of probabilistic distribution represented as phoneme posteriors {\whiteCircled{1}}. For matching, \sys{} computes the CTC loss between the phoneme posteriors and cached phoneme sequence \whiteCircled{3}. 

Cache entries are updated upon L2 cache miss: the ground truth utterance sent to cloud is tokenized and used as a ``reference'' phoneme sequence {\whiteCircled{2}}. L2 caches this phoneme sequence against a cloud provided intent. With each seen utterance in cloud, the model is finetuned {\whiteCircled{4}}.

\subsection{Designs}
\paragraph{\whiteCircled{1}~Feature extraction} 
A phoneme is a sound unit/phone spanning one or more audio frames.
We follow a common design for extracting phonetic features from raw speech. 

Consuming a sequence of uncollapsed, frame-level spectral features $f_i$ for each frame $F_i$  from L1 (step {\circled{1} in \autoref{fig:l1d}}), L2 applies two bi-directional GRU layer (a hidden size of 128), 
followed by a linear classifier and a 42 phoneme output softmax layer (41 context independent (CI) phoneme targets and an additional «sp» (blank) target). 

The output phoneme posterior $P_i$ is a sequence $X$
that represents the log probabilities or phoneme logits at each time step in the sequence; here $X =\{X_1,...,X_{42}\}$. Implementation is similar to \S \ref{l1-imp}.

\paragraph{\whiteCircled{2}~Tokenization} 
Upon L2 cache miss, the cloud provides a ``reference'' phoneme sequence $p = \{p_x, p_y,\cdots,p_z\}$ for $L$.  To do so, the cloud runtime transcribes $L$ to words and then to phonemes, 
using a standard tokenizer such as NLTK \cite{NLTK}. A special ``blank'' token is inserted between adjacent words.
The rationale is that the cloud with its large speech/language models can transcribe $L$ with low WER; the phonemes (reversed) derived from the transcript and used as cache key are therefore less error prone and closer to the ground truth. 




\paragraph{\whiteCircled{3}~Sequence matching}
For each time step T, The generated sequence of phoneme posteriors $P = (P_1,...,P_{T}) $ will be matched against all existing L2 entries $\{L_{i=0..N}\}$ which save deterministic phoneme sequences. 
Here, the match employs CTC loss: from the probabilistic $P$ and a saved sequence ($L_i$), L2 determines \underline{all} \textit{concrete} phoneme sequences (alignment) that would collapse to $L_i$. It computes the aggregated probabilities for a single alignment ($\pi$) as: $    p(\pi|P) = \prod_{t=0}^{T-1} y_{\pi_t}^t$, where $y_{\pi_t}^t$ is  the
probability of observing label $\pi_t$ at time $t$.


Then the CTC loss ($l$) is computed as the sum of the probabilities of all the valid alignments mapped onto it by $\tau$, $  p(l|P) = \sum_{\pi \in \tau^{-1} (l)} p(\pi|P)$.

The cache entry with the lowest CTC loss is chosen; if the loss is below a predefined threshold, L2 deems it as a hit and returns the associated intent.

\S \ref{sec:opt} discusses threshold determination. 

\paragraph{\whiteCircled{4}~Online finetuning} is crucial as shown in \S \ref{sec:eval} as it allows \sys{} to be personalized.
The cloud keeps a shadow copy (M) of the phoneme feature extractor that the device is currently running. 
In the training phase, given the phoneme logits (or posteriors), the model is finetuned to predict phoneme targets (ground truth). The cloud runs a forward pass of M, gets a sequence of phoneme posteriors, calculates the CTC loss between generated sequence and the reference sequence. Finally, the loss is backpropagated to update the parameters of M. Having accumulated all updates, the cloud pushes M' to the device for update. 
\vspace*{-1em}
\subsection{Implementation details}

The objective of the model is to perform a single label intent classification task. The training process involves both real and augmented inputs with batch size 16, Adam optimizer and learning rate $1e^{-4}$. 
L2 is finetuned until the CTC loss converges.
The CTC loss in PyTorch uses dynamic warp search (DTW), an optimized version of the traditional forward backward algorithm. For L2 feature extractor, hidden dimension 128 is optimal.


\section{Key optimizations}
\label{sec:opt}





\paragraph{Model Ensembling}
\label{para:audio-bucket}
A novel optimization is to specialize feature extractors to input length
-- a reasonable indicator of acoustic and lexical complexity. 
Intuitively, utterances lasting 4-5 seconds (e.g."At one pm today start the robot vacuum cleaner in kitchen") are likely queries with context; unlike short commands only lasting a few seconds (e.g. "Play music").
We hypothesize that their matching tasks could benefit from separate models finetuned on the input complexity.


Concretely, \sys{} instantiates multiple  ($K$) versions of the feature extraction models, each version finetuned on a range of input audio lengths. 
Empirically, \sys{} runs $K=3$ models, for input lengths (0,2.7] sec, [2, 4) sec, and [4,) sec respectively, also referred to as buckets. 
This multi-versioning of models can be seen as a special case of Mixture of Experts (MoE) \cite{jacobs1991adaptive}. Similar to MoE, the subdivision of predictive modeling tasks is done; division element being the input length. An expert (bucket model) is developed for each subtask. Unlike in MoE, we do not need a neural gated network for routing individual input to an expert, it is simply done by comparison against input length.


At runtime, input goes through each bucket model. After the command ends, the extracted features are taken from the bucket model corresponding to input audio length. Note that multiversioning only applies to devices; the cloud still runs a monolithic model for SOTA performance. 

L1 underperforms on short commands as it doesn’t have enough features for better inference. For such commands we bypass L1 and directly process them in L2. 
To determine the threshold for short commands -- we do a systematic exploration with cutoff values at every 0.5 intervals upto 3 seconds and observe that a cutoff of 2 second (or commands in range (0,2.7] sec) can be considered as 'short' and eligible for bypassing L1.

\paragraph{Weight Sharing} Only the weights in L2 are learnable. The acoustic feature extractor of L1 (which is also common to L2) is independent to learnable weights; thus can be shared by $K$ bucket models. 

\paragraph{Input augmentation}
To fine-tune the on-device models, a considerable difficulty is that a device may not have enough inputs, which we expect can be as few as 
5 -- 10. 
To address this difficulty, we find data augmentation vital. 

For any offloaded waveform, the cloud creates multiple augmented versions of it in order to resemble the possible variation in future inputs: 
(a) Temporal Shift -
\sys{} shifts the waveform towards either direction for $X_s$\% of the total duration, where $X_s$ conforms to uniform distribution in [-5,5]. This simulates users starting to speak before activating device or continuing to speak after recording ends.
(b) Frequency Shift: \sys{} varies the waveform frequency by $X_f$\% , where $X_f$ conforms to uniform distribution in [-10,10]. 
(c) Ambient Noise: \sys{} applies Gaussian noise at 5\% of the maximum volume in the recording. It allows the model to distinguish important phonemes from background noise. In a typical home setting a wide range of ambient noise could occur at the same time, e.g., television and phone.

Given an input, each transformation above creates five versions of it. 
    

\paragraph{In Domain training} is applicable for SLURP. Before finetuning, the base model is trained on domain specific utterances from SLURP. The purpose is to make the model familiar with the complexity of SLURP sentences. We train on 10\% of the total SLURP utterances.

\paragraph{Dynamic thresholds for CTC loss}
Intuitively, the CTC loss threshold $X$ for cache hit should correlate to the length of the input sequence $Li$.
To dynamically determine the threshold, \sys{} adopts a small, 2-layer MLP model (with hidden dimension of 64, and ReLU activation) that maps $Li \rightarrow X$. 
The model only has 193 parameters; 
during cache lookup, its inference overhead is negligible compared to other computations, e.g. CTC loss. This optimization is only needed when querying inputs in evaluation. It is not needed for learning.
We train the model offline on a held-out set loaded with 100 entries and use the MLP model to predict the threshold (that would give best results) on the fly.

\section{Evaluation}

\label{sec:eval}
We answer the following questions through experiments:
\begin{enumerate}
    \item Can \sys{} achieve competitive accuracy \& latency? 
    \item Do the key designs of \sys{} positively contribute to its performance?
    \item Is the performance of \sys{} sensitive to environments and configurations?
\end{enumerate}




\subsection{Methodology}
\label{sec:eval:method}

\begin{table}[t]
    \centering
    \includegraphics[width=.95\columnwidth]{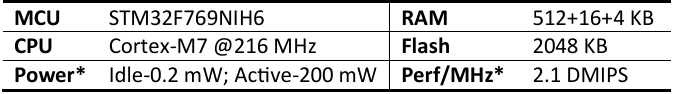}
    \caption{The test plaform. Numbers* from~\cite{st_mcu, st_mcu_2}.}
    \label{tab:tp}
    \vspace{-10mm}
\end{table}
\paragraph{Test platforms}
We implement \sys{} atop PyTorch 2.0.1 (for the cloud runtime) and X-Cube-AI 8.1.0 (for the device runtime). 
We deploy \sys{} in a low resource platform as detailed in \autoref{tab:tp}. We choose a Cortex-M7 processor as it has integrated singe-instruction multiple-data (SIMD) and multiply-accumulate (MAC) 
instructions useful for accelerating low-precision computations.

We run \sys{}'s cloud runtime on an x86 server in lab. To better estimate the cloud/device network delays in real deployment, 
we invoke Microsoft's speech service \cite{azuresdk}
with the benchmark inputs and measure end-to-end wall time. The input is sent from the US east coast and invokes data centers of the east coast.
We repeat the test on enterprise WiFi and LTE, 
and use that measurement as the offload delays in our experiments. 
Our delays are 0.29--0.34 RTF (on average: 900ms for a 3 second audio; stddev: 100ms), which largely match the cloud API delays in prior work ~\cite{xu2020cha, porcheron2018voice, potdar2021streaming} . The payload is only tens of KB hence bandwidth is not an issue. Note that, Azure speech service is used only to measure the RTT (in ms), the accuracy reported comes from the actual hardware detailed in \autoref{tab:tp}.

\paragraph{Datasets}
are summarized in Table \ref{tab:dataset-stat}: 

\noindent
(1) \underline{SLURP-C} is curated from a popular speech benchmark SLURP~\cite{bastianelli2020slurp}, which comprises lexically complex, linguistically-diverse
utterances close to daily conversations, e.g. ``please add an event to my schedule''. 
As the original SLURP waveforms were captured with varying devices and acoustic conditions,  
we construct SLURP-C as the subset recorded in the close range setting 
(2745 utterances from 74 speakers; each speaker utters 5--961 transcripts), best matching our targeted scenarios such as smart homes and wearables. 
We also consider adversarial inputs: a subset \underline{\slurpMix{}} recorded with a mix of near/far range devices and in noisy conditions (52,935 utterances from 157 speakers); 

\noindent
(2) \underline{FSC}~\cite{lugosch2019speech} comprises shorter and simpler utterances representing voice assistant commands, e.g. ``play music''; with 30K utterances from 97 speakers, covering 31 intents.
Given a speaker, a distinct transcript is uttered 1--2 times.

Note that compared to the original SLURP and FSC datasets, 
we exclude speakers that have too few (<5) utterances, as they lack enough data to warm up our cache. 
See \S \ref{sec:design} for discussion on such a situation.



\begin{table}[t]
    \centering
    \includegraphics[width=.9\columnwidth]{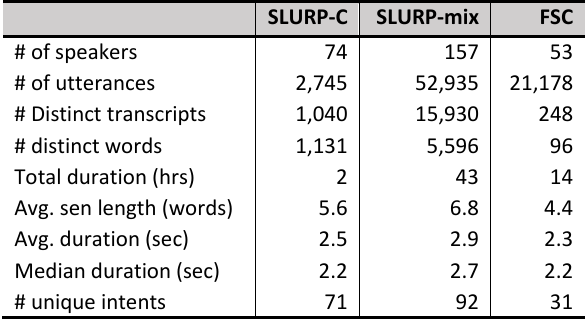}
        \caption{Datasets in experiments. See \S \ref{sec:eval:method} for details.}
\vspace{-10mm}
    \label{tab:dataset-stat}
\end{table}

\paragraph{Comparisons}
\label{baseline}
\textit{CloudOnly} offloads all inputs to the cloud, 
which runs a model that achieves the SOTA accuracy with the best efficiency, with regard to each dataset.
For SLURP-C, the cloud runs NVIDIA's Conformer-Transformer-Large \cite{ nemo_conformer_transformer_large}, pretrained on NeMo ASR-Set 3.0 and finetuned on SLURP. 
For FSC, the cloud runs an attention-based RNN sequence-to-sequence model from SpeechBrain \cite{SpeechBrain_FSC}. \textit{CloudOnly}'s accuracy is considered as gold.



\textit{OnDevice-\{S|L\}} are models that completely run on device. 
(1)~\textit{OnDevice-S} is tailored to MCUs and targets simple utterances. 
As a popular model by Fluent.AI~\cite{lugosch2019speech}, \textit{OnDevice-S}
has only 3.96M parameters and was shown to have good accuracy on FSC. 
(2)~\textit{OnDevice-L} is a compressed model that can handle complex utterances as in \slurp{}. 
It comprises a Conformer(S) \cite{gulati2020conformer} for ASR 
 and a MobileBERT \cite{sun2020mobilebert} for NLU, totalling $\sim$110M parameters (ASR:NLU=1:10). 
\textit{OnDevice-L} far exceeds an MCU's resource; 
we regard it as a reference point --  
an efficiency-optimized model that still generates reasonable accuracy on \slurp{}. 
 We confirm that further smaller models would not produce meaningful results. 
 

\textit{Ours-\{D|T|M\}\textsuperscript{+}} is our system with different combinations of optimizations (as discussed in \S \ref{sec:opt}), 
in which: 
D stands for dynamic thresholds for feature matching; 
T stands for in-domain training; 
M stands for model ensembling.

\begin{figure}[t]
\centering
\includegraphics[width=\columnwidth]{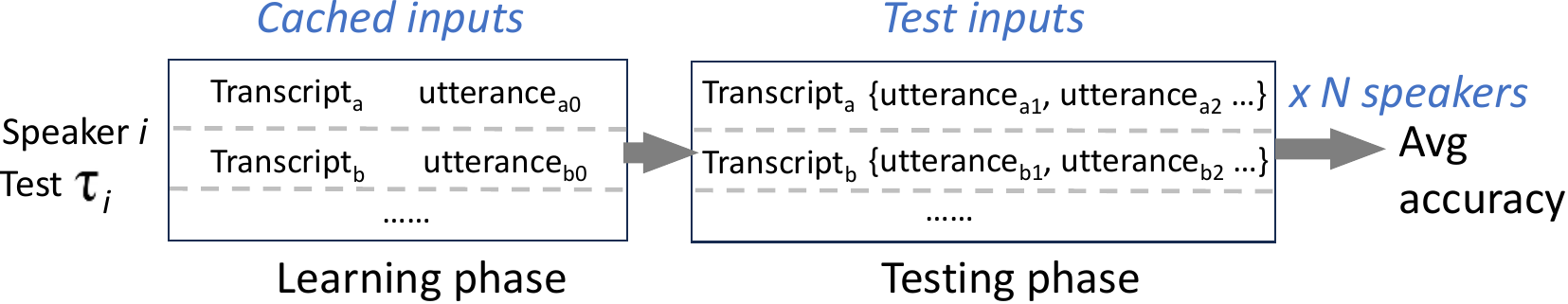}
    \caption{Benchmark settings, showing the test for one given speaker. We report the accuracy averaged over all speakers.}
    \label{fig:bs}
    \vspace{-6mm}
\end{figure}


\paragraph{Benchmark settings}
We control two factors that have high performance impact:
the number of speakers sharing a device; 
the fraction of unseen speech transcripts. 


\textit{1-speaker-100\%-seen} comprises of $74$ (\slurp{}) and $53$ (FSC) separate tests, each for a distinct speaker.
In a test $\tau_i$ for speaker $i$, 
all test inputs are from this speaker. 
We construct $\tau_i$ as follows (also see \autoref{fig:bs}). 

From the speaker $i$'s utterances corresponding to a distinct transcript, 
$\tau_i$ includes one randomly selected utterance as 
a \textit{cached input} and all other utterances as \textit{test inputs}. 
A test $\tau_i$ has two phases.
In the learning phase \sys{} processes all the \textit{cached} inputs 
(which have various transcripts), building the device cache and finetuning the feature extractors. 
In the test phase \sys{} uses the tuned feature extractors 
and processes the \textit{test} inputs. 
We report performance for the test phase: 
for each of SLURP-C and FSC, 
we aggregate the results from all the tests $\{\tau_{i=1..M}\}$, where $M$ is the number of speakers (74 for \slurp{} and 53 for FSC). 

\textit{1-speaker-k\%-seen} 
is the same as above, except that in each test $\tau_i$, 
only the transcripts of k\% of test inputs have appeared in the cached inputs. 
We experiment with k=70 and k=0 (an extreme case, no transcripts were seen). 

\textit{n-speakers-k\%-seen} 
is the same as above, except that each test $\tau$ now comprises
utterances from $n$ randomly grouped speakers. 
Speaker groups are disjoint. 
The utterances carry no speaker IDs. We report results for $n=3$ and $k=100$, 
for which we perform 24 (SLURP)  and 18 (FSC) tests. Additionally, we also test on \textit{all-speakers-100\%-seen}.


\paragraph{Metrics}
We report accuracy and latency 
averaged over all the test inputs. 
The latency is end-to-end, 
from the moment an utterance completes till the system generates a response. Real-time factor (RTF) measures how fast a speech model can process audio input and is the ratio of the processing time to audio length.
We report latency both in wall-clock time and RTF (wall time normalized by the voice length).

\subsection{End-to-end results}
\label{sec:end-to-end}


\begin{figure}[t]
\centering
\includegraphics[width=0.48\textwidth]{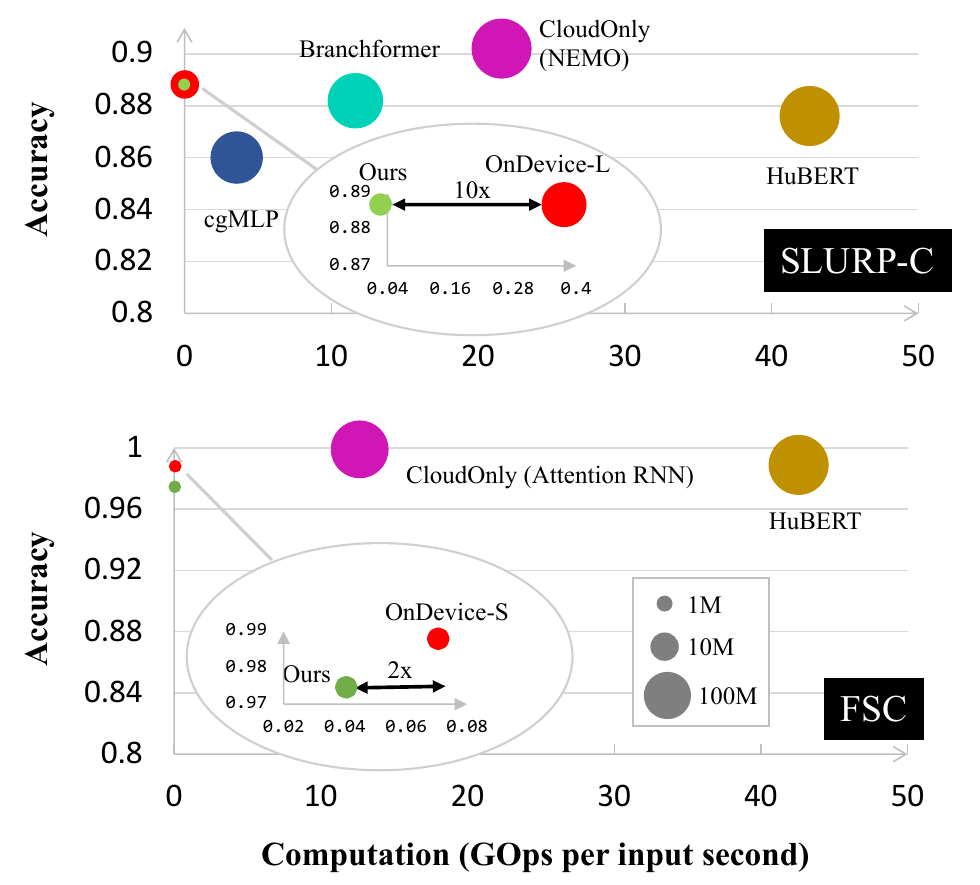}
    \caption{
    \sys{} (ours) with only 0.5M parameters and 1.8 MB model size incurs low compute per input  and delivers high accuracy. The closest other on-device model is 2x more expensive in FSC and 10x in \slurp{}.
        }
        \label{fig:fvb}
\end{figure}


\begin{table}[t]
    \centering
	\includegraphics[width=\columnwidth]{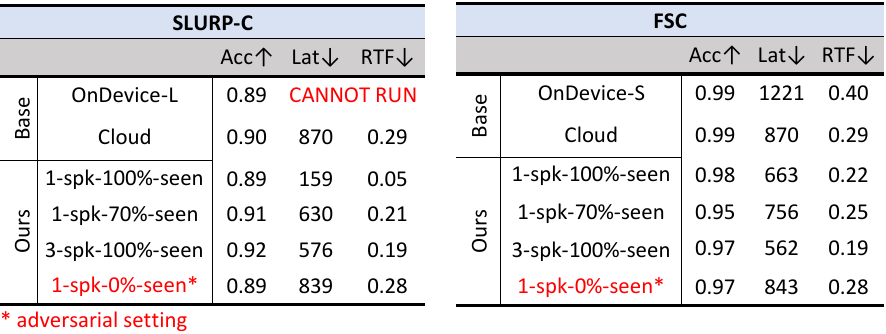}
\caption{Measured performance of our system as compared to baselines,
showing that we deliver strong accuracy while incurring much lower latency (ms in table). 
The absolute latency is for an input of 3 second. 
Note that the baselines' performance remain largely unaffected w.r.t. benchmark settings. 
}
    \vspace{-5mm}
    \label{tab:om}
\end{table}


\begin{table*}[t]
    \centering
\includegraphics[width=2\columnwidth]{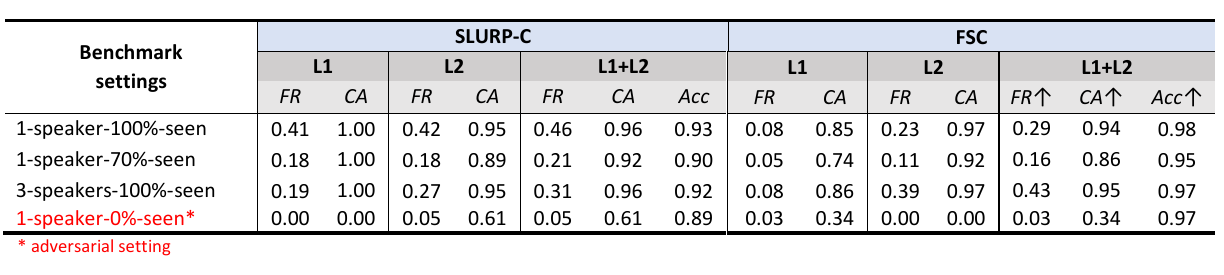}
    
    \caption{Detailed cache performance, showing the cache's efficacy and that L1/L2 complement each other. Filter rate (FR): \# of inputs hit at a cache level, normalized by \# of inputs received by that cache level. Cache accuracy (CA): of all hits at a cache level, the fraction that the cache level correctly classifies. Measurement from \textit{Ours-DM}. 
} 

    \label{tab:fsp}
\end{table*}




As shown in \autoref{fig:fvb}, our system
offers competitive latency/accuracy tradeoffs.






(1) On \textit{\slurp{}} comprising complex inputs, 
our system delivers high accuracy of 0.89
(only 0.01 lower than gold) 
while incurring 4x lower latency (159 ms vs. 870 ms). 
\textit{OnDevice-S}, which suits most MCUs, fails to generate meaningful responses: 
Its accuracy is as low as 0.06. 


Compared to \textit{OnDevice-L},
our system delivers similar accuracy at 10x less compute.
Our accuracy is on par with (or even exceed) much larger models, 
including cgMLP (0.86) \cite{rajagopal2021convolutional}, BranchFormer (0.88) \cite{peng2022branchformer}, and HuBERT (0.88) \cite{hsu2021hubert}. 
These models however require far more memory and compute than MCUs can offer.



 


(2) On \textit{FSC} which comprises simple voice inputs, our system and all baselines deliver accuracy as high as 0.98--0.99. 
Yet, our system runs much faster: its latency is 1.3x lower than \textit{CloudOnly}
and even 50\% lower (663ms vs 1221ms) as compared to \textit{Ondevice-S} of only 3.96M parameters. 

On-device compute is measured in GOps per input second using PyTorch.

\paragraph{Impact of input complexity}
Our benefit is more pronounced on complex, richer inputs as in \slurp{}.
As shown in \autoref{tab:om}, \sys{} incurs lower latency (due to higher filter rates) on \slurp{} than FSC, 
because its cache is more effective on matching longer inputs comprising richer sound features. 
By contrast: 
on these inputs, on-device SLU cannot be afforded  by MCU; 
shallow, matching-based query-by-examples (QbE) can only handle inputs as short as a few words \cite{lugosch2018donut}, 
because it lacks the personalization capability. 

\paragraph{Impact of the number of speakers}
\label{subsection:number-of-users}
Our system is robust against additional speakers sharing a device. 
Compared to the default 1-speaker setting,
\textit{3-speakers-100\%-seen} sees modestly $\sim$400ms higher latency at similar accuracy on \slurp{} as shown in \autoref{tab:om}. 
\autoref{tab:fsp} shows the breakdown results: slightly lower filter rates 
while maintaining the same level of cache accuracy. 
With additional speakers, although \sys{} loses some benefit of personalization, it is still more specialized (thus more efficient) than a generic on-deivce model trained to fit all training data. 

A generic \sys{} trained on all data from \slurp{} gives a low accuracy of 0.80, showing the significance of personalization (our design).
\vspace{-3mm}

\subsection{In-the-wild evaluation}
To thoroughly test \sys{}, we conduct a user study and report performance in the four benchmark settings.
\paragraph{Design and Data collection:}
We collect 210 audio recordings from three volunteer speakers. Among them, 126 utterances are recorded with headset (close range) and 84 without (far range). The volunteers (1 female, 2 male) are non-native speakers from diverse ethnic backgrounds (Korean, Chinese, Bengali). For the uttered transcripts,  21 unique commands were chosen from a randomly selected subset of the original SLURP dataset. The close range recording is done using a standard headset (Sony WH320) and for adversarial inputs (far-range), audio is recorded using the M1 Mac microphone. The adversarial dataset contains a mix of with and without headset recordings. 

\paragraph{Evaluation:}
While evaluating, we obtain the filter rate, cache accuracy, overall accuracy, latency and RTF in the four benchmark settings. 
We observe that \sys{} retains a similar performance on the collected recordings to that in the original experiments conducted using SLURP data, and retains a similar accuracy as before with a minor delay (end to end latency of 409ms).  For the different benchmark settings, the accuracy and latency is consistent. 
The experimental setting was \textit{Ours-M} where only the model ensembling optimization was used.
For adversarial inputs, performance is slightly inferior on our custom dataset. For a 1-spk-100\%-seen benchmark with adversarial inputs, after noise reduction, overall accuracy is 0.80 at a filter rate of 0.26 having latency of 644ms. 
The detailed benchmark results and comparison with baselines are reported in Table \ref{tab:user-study-performance}.

\begin{table}[t]
    \centering
\includegraphics[width=\columnwidth]{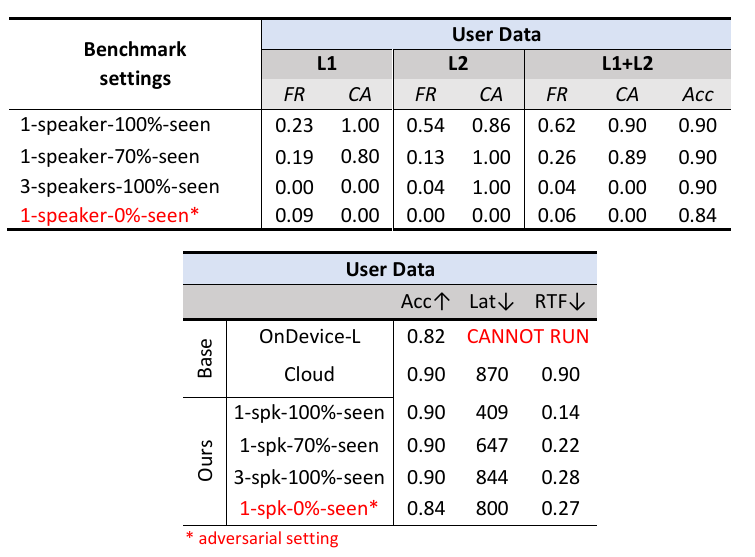}
    \caption{Our performance on real users. Absolute latency (ms in table) is for a 3-second input.
}
    \label{tab:user-study-performance}
    \vspace{-6mm}
\end{table}
\subsection{Cache efficacy}
\paragraph{Cache latency}
\sys{}'s cache is fast. 
The most expensive operation, feature extraction, is done in a streaming fashion in parallel to the voice ingestion. 
As a result, the delay for processing all but the last streaming segment (which spans around 250ms) are hidden. 
After features are extracted, the overhead of matching the features, is negligible. 
In case of L1 hit, \sys{}'s latency is 96 ms on our test platform (most of which is from SincLayer); 
in case of L1 miss and L2 hit, the latency is an additional 89 ms. 
The on-device latency is almost 5x lower than offloading to the cloud.

\paragraph{Cache accuracy} is decent as shown in Table \ref{tab:fsp}.   
(1) On inputs with known transcripts (\textit{1-speaker-100\%-seen}), 
our cache processes (i.e. filters) 46.23\% of inputs on device, avoiding offloading them.
Among such locally processed inputs, our cache's accuracy is as high as 0.96.
Between the two cache levels, 
L1 is more selective (i.e. lower filter rate) but shows higher accuracy.

(2) On inputs with unseen transcripts (e.g. \textit{1-speaker-0\%-seen}), 
the cache correctly deems almost all inputs as mismatch, offloading them to the cloud. 
Filters rates are thus very low (3-5\%). 
As a result, the overall accuracy only sees a minor drop (about 2-3\%) 
compared to the setting with all seen transcripts. 


\paragraph{Two-level caching}
Both levels complement each other and contribute to the overall performance as shown in \autoref{tab:fsp}. 
On longer inputs (\slurp{}), L1 is more effective (i.e. higher filter rates) and incurs low cost; 
yet, on short inputs, L1's simple features are often noisy. 
This is compensated by L2 with its deeper features. 
We deem both levels essential. 

\subsection{Significance of key optimizations}
\label{sec:key-op}




\begin{figure}[t]
\centering
\includegraphics[width=\columnwidth]{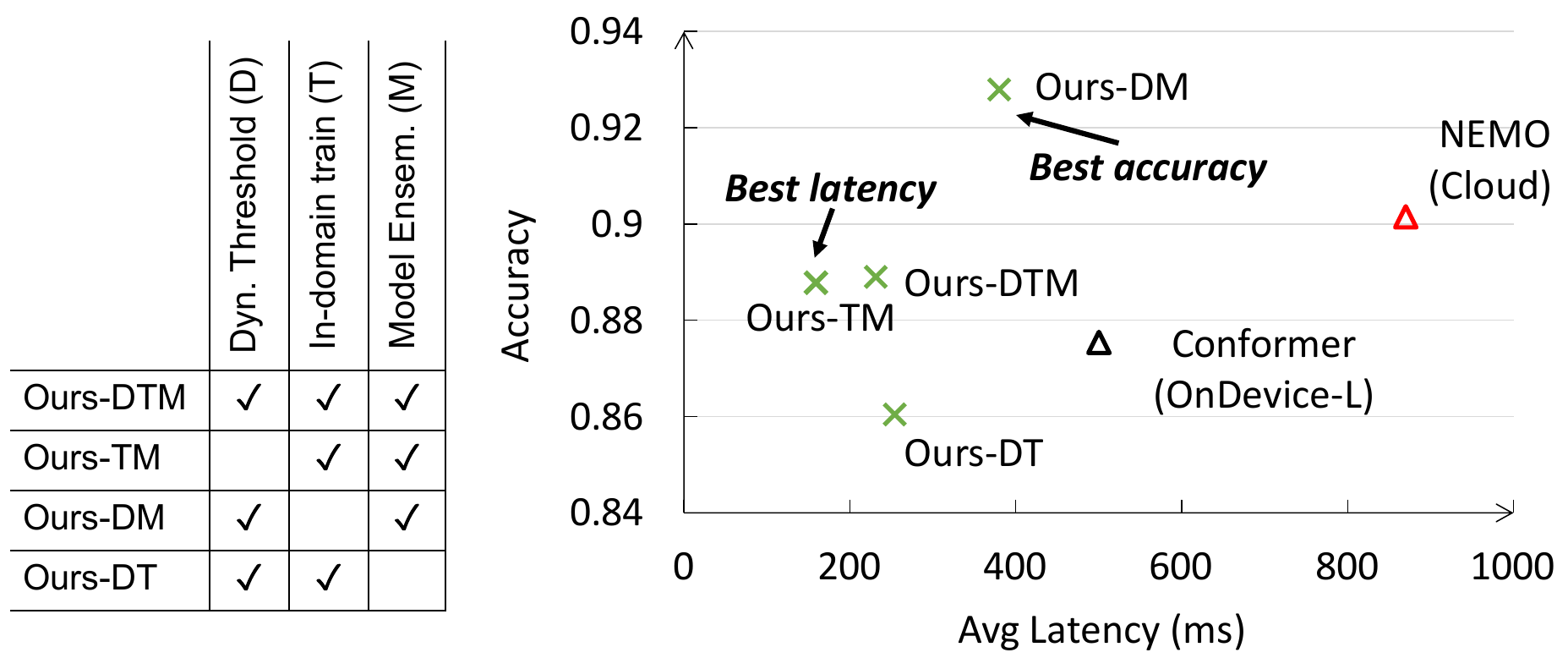}
    \caption{Our optimizations play complementary roles in the overall performance. 
    Ours-DM gives the best accuracy while Ours-TM gives the best latency.}

    \label{fig:so}
\end{figure}

\paragraph{Online learning}
\sys{} not only benefits from caching but also crucially from learning (i.e. personalization). 
Using frozen, pretrained feature extractors without online finetuning sees overall accuracy drop by 0.11 (from 0.99 to 0.88) in FSC. 
Deeper investigation shows that the cache is much less effective: 
the filter rate is as low as 2\%; 
among all the cache hits, the accuracy is reduced by 40\%.


\paragraph{Model ensembling} is vital. 
Replacing the default ensembled model
with one monolithic model results in notable accuracy drop as shown in \autoref{fig:so} (Ours-DM vs. Ours-DT). 
The reason is that using one model limits the room for input specialization, rendering caching less effective.
Further increasing the number of buckets, e.g. from 3 to 5, sees diminishing return. 



\paragraph{Dynamic thresholds}
Described in \S \ref{sec:opt}, 
the thresholds for cache matching scores hinge a tradeoff between the filter rate and the accuracy. 
We compared two approaches: 
(1) predefined thresholds per input bucket, and (2) 
predicting the threshold based on an input's duration. 
As shown in \autoref{fig:so}, dynamic threshold when complemented with model ensembling (Ours-DM) sees notable accuracy improvement  compared to the one without dynamic threshold (Ours-TM).

\begin{table}[t]
    \centering
\includegraphics[width=0.63\columnwidth]{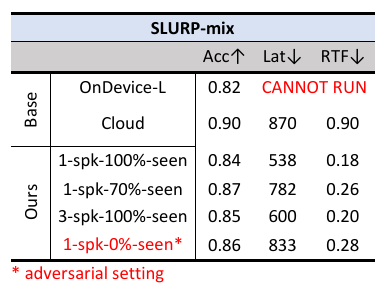}
    \caption{Our performance on \slurpMix{} (adversarial workloads). 
    Absolute latency (ms in table) is for a 3-second input. 
}
    \label{tab:SLURP-all-measured}
    \vspace{-6mm}
\end{table}

\paragraph{Robustness against acoustic environments}
Even on adversarial input conditions, 
\sys{} achieves our goal of faster inference with minor reduction in accuracy. 
For a mix of near and far range inputs in \slurpMix{}, latency increase is only $\sim$300 ms (\slurpMix{} 538 ms vs \slurp{} 159 ms) at 0.84 accuracy, shown in \autoref{tab:SLURP-all-measured}. This performance drop of 0.05 is acceptable given the adversarial input condition.
\begin{table}[t]
    \centering
    \includegraphics[width=\columnwidth]{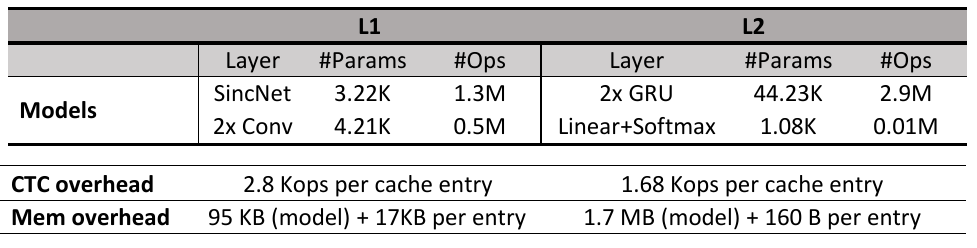}
    
    \caption{Overhead breakdown, showing our light computation and low memory footprint. Model ops is per input; CTC overhead is per cache entry.}
  
	\label{tab:comp}
   \vspace{-1cm}
\end{table}
\subsection{Overhead analysis}
\label{overhead-analysis}
\paragraph{Computational Complexity}
\label{para:computational-complexity}
 \autoref{tab:comp} provides the computational complexity of \sys{} measured as the number of multiplication and addition operations, or MOps per input second. We find that -
  
(1) Overall, \sys{} includes low computation complexity on device.
As a reference, deep (SOTA) SLU cloud models are often 12 --- 60 GOps/s (200-1000 orders higher). 

(2) Hardware Measurements: Between the two levels, for every 10 frames, L1  (SincNet + 2x CNN) performs 1.8 MOps at 96 ms latency; L2 (2x GRU) performs an additional 2.9 MOps at 89 ms as measured on device; on-device processing latency totalling at 185ms. L1 incurs 37.93\% lower MOps than L2 due to L1's emphasis on local features and its exclusion of temporal dependency (using only Convolution).

(3) The cost of feature extraction dominates during inference. As a comparison - per entry cache match cost of \sys{} is only 2.80 KOps in L1 and 1.68 KOps in L2 (shown in \autoref{tab:comp}). In theory, complexity of CTC is $O(v^T)$, $v$ being size of vocabulary, $T$ timesteps. Using DP in CTC (norm) reduces complexity from order of $T$ to a polynomial function $(v \times T)$. For $n$ entries in cache, each match requires $v \times T \times n$ computations. From workload study (median 32 commands/week \cite{garg2020he}, median 4.1 commands/day \cite{bentley2018understanding}), we expect $n$ to be typically tens, no more than a few hundreds.
With such a size, the delay in cache lookup is in the order of $n$. After model ensembling, the search complexity is reduced by one-third.


\paragraph{Memory footprint} 
\autoref{tab:comp} shows the memory footprint of \sys{}. The size of a single \sys{} model is about 1.8 MB before compression. As a standard practice, we apply 8-bit dynamic quantization and further shrink the model size to 0.62 MB.
The deployed model is made up of parameters (weights and biases) and cached values (key-value pairs). L1 cache entries are larger since they store KxN cluster centers. Even after multiversioning, the deployed \sys{} model occupies less than 1.9 MB of memory.

\paragraph{Energy and Power Consumption}
\sys{} is not an always-on system, hence power consumption is low. We only need the KWS system (a separate module) of the device to be always-on. In stop mode \sys{} consumes only 100uA power (or ~0.2mW at 1.7V) and in active mode 118mA power (or 200.6 mW at 1.7V)\footnote{\href{https://www.st.com/en/microcontrollers-microprocessors/stm32f7x9.html}{www.st.com/en/microcontrollers-microprocessors/stm32f7x9.html}}. (1) When inference is done fully on-device – given an end-to-end latency of 185ms and following the power-energy equation (E=Pt), the total energy consumption is only 37 mJ per utterance (2) when there is a need to offload – an additional power is consumed for transmission, the delay is 900ms on average for a 3 second audio with a standard deviation of 100ms as described in \autoref{sec:eval:method}. By the first order approximation and taking on device execution delay, the total energy consumption is 180 mJ per utterance.

\vspace{-2mm}
\section{Related work}
\label{sec:related-work}

%

\paragraph{Spoken Language Understanding (SLU)} is discussed in \S \ref{sec:slu}. 
Existent SLU architectures employ convolutional \cite{mhiri2020low, tian2020improving} or recurrent \cite{palogiannidi2020end, liu2016joint, potdar2021streaming, lugosch2019speech} networks and perform compression to deploy on end device.

\paragraph{Keyword spotting (KWS)} 
is defined as the task of identifying a predefined keyword (such as ``hey google``) in audio streams \cite{lopez2021deep, vinyals2014chasing}.
A small footprint KWS model does not scale well to large vocabularies and cannot perform well for longer commands. The reverse:  SLU capability on KWS tasks is verified by \cite{potdar2021streaming} - their E2E SLU model gives acceptable (if not better) accuracy on standard KWS tasks. 



\paragraph{Query by Example (QbE)} is a template matching technique used for spoken term detection (STD) or KWS \cite{hazen2009query, lugosch2018donut}, the objective of which is to locate audio snippets containing keywords (enrollments) within a larger audio stream. But this technique is not well-suited for IC as the users need to explicitly define the "queries". 
Our feature extractor is significant since we can fine tune it (learnable) unlike QbE.
\paragraph{Intent Classification (IC)} is discussed in \S \ref{sec:slu}. 
IC is much more challenging (more words, requires contextual understanding) compared to KWS.

Prior work introduces a monolithic model, deployed once, for all (possible) users \cite{mun2020accelerating, he2023can, lugosch2019speech} but this generalization power is an overkill \cite{arora2023universlu}. 
\cite{myer2018efficient} uses a time delay NN architecture for KWS, caches intermediate results and skips frame to reduce compute. \cite{he2023can} incorporates prompting and in-context learning for IC using large pretrained language models. 

\cite{vacher2015speech} delineates the correlation between speech recognition performance and training size; an acceptable performance requires > 100 training data. As discussed in \S \ref{sec:design}, our model is robust against this challenge. \cite{long2023ai} mentions exploring the opportunity to focus on short, complex commands with multiple elements instead of extended conversations. We dwell in the middle and focus on longer, complex but single intent commands. \cite{xu2020cha} introduces an intermediate caching framework (or hub) in between the edge and cloud for IC.
\cite{mcgraw2016personalized} employs a moderately small LSTM network for  large vocabulary speech recognition.
Multiple models employ content prefetching for service acceleration \cite{mun2020accelerating, schwarz2023personalized}. 
A concurrent work optimizes SLU for Armv8 SoCs in the local/cloud setting \cite{anonymous2023speechhybrid}. Unlike Armv8 SoCs that can run a \textit{complete} SLU engine, microcontrollers can only run an inference \textit{cache} (our unique design). These two projects do not depend on each other. Their contributions are orthogonal. 

Additionally, commercial off the shelf speech-to-intent engines such as rhino from PicoVoice \cite{picovoice_rhino}, Wio Terminal \cite{WIO_terminal} etc. cannot scale to the performance of \sys{}. 

\section{Conclusions}
In this work, we propose a novel, hierarchical, and learning cache for end-to-end SLU inference. It leverages the benefits of both on-device and cloud infrastructure. \sys{} exploits temporal locality of voice commands and performs SLU at 80\% of the cost with comparable accuracy. It is capable to execute on an MCU with just 2 MB of memory. Moreover, we implement a series of novel optimizations to increase performance and adapt to resource requirements on tiny devices.

\section*{ACKNOWLEDGMENT}

The authors were supported in part by NSF awards \#2128725,
\#1919197 and \#2106893. The authors thank the anonymous reviewers for their insightful feedback.

%




\bibliographystyle{plain}
\bibliography{bib/main.bib}

\begin{thebibliography}{10}

\bibitem{nemo_conformer_transformer_large}
\href{https://catalog.ngc.nvidia.com/orgs/nvidia/teams/nemo/models/slu_conformer_transformer_large_slurp}{\color{black}nvidia/teams/nemo/models/slu\_conformer\_transformer\_large\_slurp}.

\bibitem{azuresdk}
\href{https://github.com/Azure-Samples/cognitive-services-speech-sdk}{https://github.com/Azure-Samples/cognitive-services-speech-sdk}.

\bibitem{picovoice_rhino}
\href{https://github.com/Picovoice/rhino}{https://github.com/Picovoice/rhino}.

\bibitem{SpeechBrain_FSC}
\href{https://huggingface.co/speechbrain/slu-direct-fluent-speech-commands-librispeech-asr
  }{https://huggingface.co/speechbrain/slu-direct-fluent-speech-commands-librispeech-asr
  }.

\bibitem{stat}
\href{https://www.demandsage.com/voice-search-statistics/}{https://www.demandsage.com/voice-search-statistics/}.

\bibitem{NLTK}
\href{https://www.nltk.org/api/nltk.tokenize.html}{
  https://www.nltk.org/api/nltk.tokenize.html}.

\bibitem{qualcomm-report}
\href{https://www.qualcomm.com/content/dam/qcomm-martech/dm-assets/documents/Whitepaper-The-future-of-AI-is-hybrid-Part-2-Qualcomm-is-uniquely-positioned-to-scale-hybrid-AI.pdf}{https://www.qualcomm.com/content/dam/qcomm-martech/dm-assets/documents/Whitepaper-The-future-of-AI-is-hybrid-Part-2-Qualcomm-is-uniquely-positioned-to-scale-hybrid-AI.pdf}.

\bibitem{st_mcu}
\href{https://www.st.com/content/st_com/en/arm-32-bit-microcontrollers/arm-cortex-m7.html}{www.st.com/content/st\_com/en/arm-32-bit-microcontrollers/arm-cortex-m7.html}.

\bibitem{st_mcu_2}
\href{https://www.st.com/en/microcontrollers-microprocessors/stm32f769ni.html}{www.st.com/en/microcontrollers-microprocessors/stm32f769ni.html}.

\bibitem{WIO_terminal}
{\href{https://github.com/AIWintermuteAI/Speech-to-Intent-Micro}{https://github.com/AIWintermuteAI/Speech-to-Intent-Micro}}.

\bibitem{microsoft2023azurehybridbenefit}
Azure hybrid benefit.
\newblock 2023.
\newblock Accessed: 2023-11-18.

\bibitem{qualcomm2023hybridcost}
Generative ai trends by the numbers: Costs, resources, parameters and more.
\newblock 2023.
\newblock Accessed: 2023-7-26.

\bibitem{ammari2019music}
Tawfiq Ammari, Jofish Kaye, Janice~Y Tsai, and Frank Bentley.
\newblock Music, search, and iot: How people (really) use voice assistants.
\newblock {\em ACM Trans. Comput. Hum. Interact.}, 26(3):17--1, 2019.

\bibitem{anonymous2023speechhybrid}
Anonymous.
\newblock Turbocharge deep speech understanding on the edge, 2024.
\newblock (Reviewers: The paper was shared with the PC chairs).

\bibitem{arik2017convolutional}
Sercan~O Arik, Markus Kliegl, Rewon Child, Joel Hestness, Andrew Gibiansky,
  Chris Fougner, Ryan Prenger, and Adam Coates.
\newblock Convolutional recurrent neural networks for small-footprint keyword
  spotting.
\newblock {\em arXiv preprint arXiv:1703.05390}, 2017.

\bibitem{arora2023universlu}
Siddhant Arora, Hayato Futami, Jee-weon Jung, Yifan Peng, Roshan Sharma, Yosuke
  Kashiwagi, Emiru Tsunoo, and Shinji Watanabe.
\newblock Universlu: Universal spoken language understanding for diverse
  classification and sequence generation tasks with a single network.
\newblock {\em arXiv preprint arXiv:2310.02973}, 2023.

\bibitem{baevski2021unsupervised}
Alexei Baevski, Wei-Ning Hsu, Alexis Conneau, and Michael Auli.
\newblock Unsupervised speech recognition.
\newblock {\em Advances in Neural Information Processing Systems},
  34:27826--27839, 2021.

\bibitem{bastianelli2020slurp}
Emanuele Bastianelli, Andrea Vanzo, Pawel Swietojanski, and Verena Rieser.
\newblock Slurp: A spoken language understanding resource package.
\newblock {\em arXiv preprint arXiv:2011.13205}, 2020.

\bibitem{bentley2018understanding}
Frank Bentley, Chris Luvogt, Max Silverman, Rushani Wirasinghe, Brooke White,
  and Danielle Lottridge.
\newblock Understanding the long-term use of smart speaker assistants.
\newblock {\em Proceedings of the ACM on Interactive, Mobile, Wearable and
  Ubiquitous Technologies}, 2(3):1--24, 2018.

\bibitem{chen2014small}
Guoguo Chen, Carolina Parada, and Georg Heigold.
\newblock Small-footprint keyword spotting using deep neural networks.
\newblock In {\em 2014 IEEE international conference on acoustics, speech and
  signal processing (ICASSP)}, pages 4087--4091. IEEE, 2014.

\bibitem{chen2015query}
Guoguo Chen, Carolina Parada, and Tara~N Sainath.
\newblock Query-by-example keyword spotting using long short-term memory
  networks.
\newblock In {\em 2015 IEEE International Conference on Acoustics, Speech and
  Signal Processing (ICASSP)}, pages 5236--5240. IEEE, 2015.

\bibitem{chen2019bert}
Qian Chen, Zhu Zhuo, and Wen Wang.
\newblock Bert for joint intent classification and slot filling.
\newblock {\em arXiv preprint arXiv:1902.10909}, 2019.

\bibitem{chung2020splat}
Yu-An Chung, Chenguang Zhu, and Michael Zeng.
\newblock Splat: Speech-language joint pre-training for spoken language
  understanding.
\newblock {\em arXiv preprint arXiv:2010.02295}, 2020.

\bibitem{coates2012learning}
Adam Coates and Andrew~Y Ng.
\newblock Learning feature representations with k-means.
\newblock In {\em Neural Networks: Tricks of the Trade: Second Edition}, pages
  561--580. Springer, 2012.

\bibitem{coucke2018snips}
Alice Coucke, Alaa Saade, Adrien Ball, Th{\'e}odore Bluche, Alexandre Caulier,
  David Leroy, Cl{\'e}ment Doumouro, Thibault Gisselbrecht, Francesco
  Caltagirone, Thibaut Lavril, et~al.
\newblock Snips voice platform: an embedded spoken language understanding
  system for private-by-design voice interfaces.
\newblock {\em arXiv preprint arXiv:1805.10190}, 2018.

\bibitem{coulthard2004author}
Malcolm Coulthard.
\newblock Author identification, idiolect, and linguistic uniqueness.
\newblock {\em Applied linguistics}, 25(4):431--447, 2004.

\bibitem{dai2017very}
Wei Dai, Chia Dai, Shuhui Qu, Juncheng Li, and Samarjit Das.
\newblock Very deep convolutional neural networks for raw waveforms.
\newblock In {\em 2017 IEEE international conference on acoustics, speech and
  signal processing (ICASSP)}, pages 421--425. IEEE, 2017.

\bibitem{denisov2020pretrained}
Pavel Denisov and Ngoc~Thang Vu.
\newblock Pretrained semantic speech embeddings for end-to-end spoken language
  understanding via cross-modal teacher-student learning.
\newblock {\em arXiv preprint arXiv:2007.01836}, 2020.

\bibitem{garg2020he}
Radhika Garg and Subhasree Sengupta.
\newblock He is just like me: a study of the long-term use of smart speakers by
  parents and children.
\newblock {\em Proceedings of the ACM on Interactive, Mobile, Wearable and
  Ubiquitous Technologies}, 4(1):1--24, 2020.

\bibitem{graves2006connectionist}
Alex Graves, Santiago Fern{\'a}ndez, Faustino Gomez, and J{\"u}rgen
  Schmidhuber.
\newblock Connectionist temporal classification: labelling unsegmented sequence
  data with recurrent neural networks.
\newblock In {\em Proceedings of the 23rd international conference on Machine
  learning}, pages 369--376, 2006.

\bibitem{gulati2020conformer}
Anmol Gulati, James Qin, Chung-Cheng Chiu, Niki Parmar, Yu~Zhang, Jiahui Yu,
  Wei Han, Shibo Wang, Zhengdong Zhang, Yonghui Wu, et~al.
\newblock Conformer: Convolution-augmented transformer for speech recognition.
\newblock {\em arXiv preprint arXiv:2005.08100}, 2020.

\bibitem{haghani2018audio}
Parisa Haghani, Arun Narayanan, Michiel Bacchiani, Galen Chuang, Neeraj Gaur,
  Pedro Moreno, Rohit Prabhavalkar, Zhongdi Qu, and Austin Waters.
\newblock From audio to semantics: Approaches to end-to-end spoken language
  understanding.
\newblock In {\em 2018 IEEE Spoken Language Technology Workshop (SLT)}, pages
  720--726. IEEE, 2018.

\bibitem{hazen2009query}
Timothy~J Hazen, Wade Shen, and Christopher White.
\newblock Query-by-example spoken term detection using phonetic posteriorgram
  templates.
\newblock In {\em 2009 IEEE Workshop on Automatic Speech Recognition \&
  Understanding}, pages 421--426. IEEE, 2009.

\bibitem{he2023can}
Mutian He and Philip~N Garner.
\newblock Can chatgpt detect intent? evaluating large language models for
  spoken language understanding.
\newblock {\em arXiv preprint arXiv:2305.13512}, 2023.

\bibitem{hsu2021hubert}
Wei-Ning Hsu, Benjamin Bolte, Yao-Hung~Hubert Tsai, Kushal Lakhotia, Ruslan
  Salakhutdinov, and Abdelrahman Mohamed.
\newblock Hubert: Self-supervised speech representation learning by masked
  prediction of hidden units.
\newblock {\em IEEE/ACM Transactions on Audio, Speech, and Language
  Processing}, 29:3451--3460, 2021.

\bibitem{huang2023leveraging}
He~Huang, Jagadeesh Balam, and Boris Ginsburg.
\newblock Leveraging pretrained asr encoders for effective and efficient
  end-to-end speech intent classification and slot filling.
\newblock {\em arXiv preprint arXiv:2307.07057}, 2023.

\bibitem{jacobs1991adaptive}
Robert~A Jacobs, Michael~I Jordan, Steven~J Nowlan, and Geoffrey~E Hinton.
\newblock Adaptive mixtures of local experts.
\newblock {\em Neural computation}, 3(1):79--87, 1991.

\bibitem{johnson19942q}
Theodore Johnson, Dennis Shasha, et~al.
\newblock 2q: a low overhead high performance bu er management replacement
  algorithm.
\newblock In {\em Proceedings of the 20th International Conference on Very
  Large Data Bases}, pages 439--450. Citeseer, 1994.

\bibitem{Jurafsky2009}
Dan Jurafsky and James~H. Martin.
\newblock {\em Speech and language processing : an introduction to natural
  language processing, computational linguistics, and speech recognition}.
\newblock Pearson Prentice Hall, Upper Saddle River, N.J., 2009.

\bibitem{kim2019query}
Byeonggeun Kim, Mingu Lee, Jinkyu Lee, Yeonseok Kim, and Kyuwoong Hwang.
\newblock Query-by-example on-device keyword spotting.
\newblock In {\em 2019 IEEE Automatic Speech Recognition and Understanding
  Workshop (ASRU)}, pages 532--538. IEEE, 2019.

\bibitem{kohrs2016delays}
Christin Kohrs, Nicole Angenstein, and Andr{\'e} Brechmann.
\newblock Delays in human-computer interaction and their effects on brain
  activity.
\newblock {\em PloS one}, 11(1):e0146250, 2016.

\bibitem{lim2021multispeaker}
Yongwan Lim, Asterios Toutios, Yannick Bliesener, Ye~Tian, Sajan~Goud Lingala,
  Colin Vaz, Tanner Sorensen, Miran Oh, Sarah Harper, Weiyi Chen, et~al.
\newblock A multispeaker dataset of raw and reconstructed speech production
  real-time mri video and 3d volumetric images.
\newblock {\em Scientific data}, 8(1):187, 2021.

\bibitem{liu2016joint}
Bing Liu and Ian Lane.
\newblock Joint online spoken language understanding and language modeling with
  recurrent neural networks.
\newblock {\em arXiv preprint arXiv:1609.01462}, 2016.

\bibitem{lloyd1982least}
Stuart Lloyd.
\newblock Least squares quantization in pcm.
\newblock {\em IEEE transactions on information theory}, 28(2):129--137, 1982.

\bibitem{long2023ai}
Tao Long and Lydia Chilton.
\newblock Ai and design opportunities for smart speakers.
\newblock 2023.

\bibitem{lopez2021deep}
Iv{\'a}n L{\'o}pez-Espejo, Zheng-Hua Tan, John~HL Hansen, and Jesper Jensen.
\newblock Deep spoken keyword spotting: An overview.
\newblock {\em IEEE Access}, 10:4169--4199, 2021.

\bibitem{lugosch2018donut}
Loren Lugosch, Samuel Myer, and Vikrant~Singh Tomar.
\newblock Donut: Ctc-based query-by-example keyword spotting.
\newblock {\em arXiv preprint arXiv:1811.10736}, 2018.

\bibitem{lugosch2019speech}
Loren Lugosch, Mirco Ravanelli, Patrick Ignoto, Vikrant~Singh Tomar, and Yoshua
  Bengio.
\newblock Speech model pre-training for end-to-end spoken language
  understanding.
\newblock {\em arXiv preprint arXiv:1904.03670}, 2019.

\bibitem{mcgraw2016personalized}
Ian McGraw, Rohit Prabhavalkar, Raziel Alvarez, Montse~Gonzalez Arenas,
  Kanishka Rao, David Rybach, Ouais Alsharif, Ha{\c{s}}im Sak, Alexander
  Gruenstein, Fran{\c{c}}oise Beaufays, et~al.
\newblock Personalized speech recognition on mobile devices.
\newblock In {\em 2016 IEEE International Conference on Acoustics, Speech and
  Signal Processing (ICASSP)}, pages 5955--5959. IEEE, 2016.

\bibitem{mctear2016conversational}
Michael~Frederick McTear, Zoraida Callejas, and David Griol.
\newblock {\em The conversational interface}, volume~6.
\newblock Springer, 2016.

\bibitem{mhiri2020low}
Mohamed Mhiri, Samuel Myer, and Vikrant~Singh Tomar.
\newblock A low latency asr-free end to end spoken language understanding
  system.
\newblock {\em arXiv preprint arXiv:2011.04884}, 2020.

\bibitem{mittermaier2020small}
Simon Mittermaier, Ludwig K{\"u}rzinger, Bernd Waschneck, and Gerhard Rigoll.
\newblock Small-footprint keyword spotting on raw audio data with
  sinc-convolutions.
\newblock In {\em ICASSP 2020-2020 IEEE International Conference on Acoustics,
  Speech and Signal Processing (ICASSP)}, pages 7454--7458. IEEE, 2020.

\bibitem{mtibaa2013towards}
Abderrahmen Mtibaa, Khaled~A Harras, and Afnan Fahim.
\newblock Towards computational offloading in mobile device clouds.
\newblock In {\em 2013 IEEE 5th international conference on cloud computing
  technology and science}, volume~1, pages 331--338. IEEE, 2013.

\bibitem{mun2020accelerating}
Hyunsu Mun and Youngseok Lee.
\newblock Accelerating smart speaker service with content prefetching and local
  control.
\newblock In {\em 2020 IEEE 17th Annual Consumer Communications \& Networking
  Conference (CCNC)}, pages 1--6. IEEE, 2020.

\bibitem{myer2018efficient}
Samuel Myer and Vikrant~Singh Tomar.
\newblock Efficient keyword spotting using time delay neural networks.
\newblock {\em arXiv preprint arXiv:1807.04353}, 2018.

\bibitem{palaz2015analysis}
Dimitri Palaz, Ronan Collobert, et~al.
\newblock Analysis of cnn-based speech recognition system using raw speech as
  input.
\newblock Technical report, Idiap, 2015.

\bibitem{palogiannidi2020end}
Elisavet Palogiannidi, Ioannis Gkinis, George Mastrapas, Petr Mizera, and
  Themos Stafylakis.
\newblock End-to-end architectures for asr-free spoken language understanding.
\newblock In {\em ICASSP 2020-2020 IEEE International Conference on Acoustics,
  Speech and Signal Processing (ICASSP)}, pages 7974--7978. IEEE, 2020.

\bibitem{peng2022branchformer}
Yifan Peng, Siddharth Dalmia, Ian Lane, and Shinji Watanabe.
\newblock Branchformer: Parallel mlp-attention architectures to capture local
  and global context for speech recognition and understanding.
\newblock In {\em International Conference on Machine Learning}, pages
  17627--17643. PMLR, 2022.

\bibitem{porcheron2018voice}
Martin Porcheron, Joel~E Fischer, Stuart Reeves, and Sarah Sharples.
\newblock Voice interfaces in everyday life.
\newblock In {\em proceedings of the 2018 CHI conference on human factors in
  computing systems}, pages 1--12, 2018.

\bibitem{potdar2021streaming}
Nihal Potdar, Anderson~R Avila, Chao Xing, Dong Wang, Yiran Cao, and Xiao Chen.
\newblock A streaming end-to-end framework for spoken language understanding.
\newblock {\em arXiv preprint arXiv:2105.10042}, 2021.

\bibitem{qin2019stack}
Libo Qin, Wanxiang Che, Yangming Li, Haoyang Wen, and Ting Liu.
\newblock A stack-propagation framework with token-level intent detection for
  spoken language understanding.
\newblock {\em arXiv preprint arXiv:1909.02188}, 2019.

\bibitem{qin2021survey}
Libo Qin, Tianbao Xie, Wanxiang Che, and Ting Liu.
\newblock A survey on spoken language understanding: Recent advances and new
  frontiers.
\newblock {\em arXiv preprint arXiv:2103.03095}, 2021.

\bibitem{radfar2021fans}
Martin Radfar, Athanasios Mouchtaris, Siegfried Kunzmann, and Ariya Rastrow.
\newblock Fans: Fusing asr and nlu for on-device slu.
\newblock {\em arXiv preprint arXiv:2111.00400}, 2021.

\bibitem{rajagopal2021convolutional}
A~Rajagopal and V~Nirmala.
\newblock Convolutional gated mlp: Combining convolutions \& gmlp.
\newblock {\em arXiv preprint arXiv:2111.03940}, 2021.

\bibitem{ravanelli2018speaker}
Mirco Ravanelli and Yoshua Bengio.
\newblock Speaker recognition from raw waveform with sincnet.
\newblock In {\em 2018 IEEE spoken language technology workshop (SLT)}, pages
  1021--1028. IEEE, 2018.

\bibitem{schwarz2023personalized}
Andreas Schwarz, Di~He, Maarten Van~Segbroeck, Mohammed Hethnawi, and Ariya
  Rastrow.
\newblock Personalized predictive asr for latency reduction in voice
  assistants.
\newblock {\em arXiv preprint arXiv:2305.13794}, 2023.

\bibitem{sciutohey}
Alex Sciuto, Arnita Saini, Jodi Forlizzi, and Jason~I Hong.
\newblock “hey alexa, what's up?”: studies of in-home conversational agent
  usage,”.
\newblock In {\em Proceedings of the DIS}.

\bibitem{serdyuk2018towards}
Dmitriy Serdyuk, Yongqiang Wang, Christian Fuegen, Anuj Kumar, Baiyang Liu, and
  Yoshua Bengio.
\newblock Towards end-to-end spoken language understanding.
\newblock In {\em 2018 IEEE International Conference on Acoustics, Speech and
  Signal Processing (ICASSP)}, pages 5754--5758. IEEE, 2018.

\bibitem{sun2020mobilebert}
Zhiqing Sun, Hongkun Yu, Xiaodan Song, Renjie Liu, Yiming Yang, and Denny Zhou.
\newblock Mobilebert: a compact task-agnostic bert for resource-limited
  devices.
\newblock {\em arXiv preprint arXiv:2004.02984}, 2020.

\bibitem{tian2020improving}
Yusheng Tian and Philip~John Gorinski.
\newblock Improving end-to-end speech-to-intent classification with reptile.
\newblock {\em arXiv preprint arXiv:2008.01994}, 2020.

\bibitem{vacher2015speech}
Michel Vacher, Benjamin Lecouteux, Javier~Serrano Romero, Moez Ajili,
  Fran{\c{c}}ois Portet, and Solange Rossato.
\newblock Speech and speaker recognition for home automation: Preliminary
  results.
\newblock In {\em 2015 International Conference on Speech Technology and
  Human-Computer Dialogue (SpeD)}, pages 1--10. IEEE, 2015.

\bibitem{vinyals2014chasing}
Oriol Vinyals and Steven Wegmann.
\newblock Chasing the metric: Smoothing learning algorithms for keyword
  detection.
\newblock In {\em 2014 IEEE International Conference on Acoustics, Speech and
  Signal Processing (ICASSP)}, pages 3301--3305. IEEE, 2014.

\bibitem{vipperla2020learning}
Ravichander Vipperla, Samin Ishtiaq, Rui Li, Sourav Bhattacharya, Ilias
  Leontiadis, and Nicholas~D Lane.
\newblock Learning to listen... on-device: Present and future perspectives of
  on-device asr.
\newblock {\em GetMobile: Mobile Computing and Communications}, 23(4):5--9,
  2020.

\bibitem{9355621}
Lanyu Xu, Arun Iyengar, and Weisong Shi.
\newblock Cha: A caching framework for home-based voice assistant systems.
\newblock In {\em 2020 IEEE/ACM Symposium on Edge Computing (SEC)}, pages
  293--306, 2020.

\bibitem{xu2020cha}
Lanyu Xu, Arun Iyengar, and Weisong Shi.
\newblock Cha: A caching framework for home-based voice assistant systems.
\newblock In {\em 2020 IEEE/ACM Symposium on Edge Computing (SEC)}, pages
  293--306. IEEE, 2020.

\bibitem{zhang2017hello}
Yundong Zhang, Naveen Suda, Liangzhen Lai, and Vikas Chandra.
\newblock Hello edge: Keyword spotting on microcontrollers.
\newblock {\em arXiv preprint arXiv:1711.07128}, 2017.

\bibitem{zhu2017encoder}
Su~Zhu and Kai Yu.
\newblock Encoder-decoder with focus-mechanism for sequence labelling based
  spoken language understanding.
\newblock In {\em 2017 IEEE International Conference on Acoustics, Speech and
  Signal Processing (ICASSP)}, pages 5675--5679. IEEE, 2017.

\end{thebibliography}
\end{document}